\documentclass[twocolumn]{aastex62}

\usepackage{xcolor}
\newbox{\bigpicturebox}

\definecolor{mygreen}{RGB}{56,150,9}
\definecolor{myteal}{RGB}{252,3,140}

\graphicspath{{./}{figures/}}

\received{September 26, 2020}
\revised{\today}
\accepted{June 8, 2021}

\submitjournal{ApJ}

\shorttitle{AT~2019krl }
\shortauthors{Andrews et al.}

\begin{document}

\title{The Blue Supergiant Progenitor of the Supernova Imposter AT~2019krl}

\correspondingauthor{Jennifer E. Andrews}
\email{jandrews@gemini.edu}

\author[0000-0003-0123-0062]{Jennifer E. Andrews}
\affil{Steward Observatory, University of Arizona, 933 North Cherry Avenue, Tucson, AZ 85721-0065, USA}

\author[0000-0001-5754-4007]{Jacob E. Jencson}
\affil{Steward Observatory, University of Arizona, 933 North Cherry Avenue, Tucson, AZ 85721-0065, USA}

\author[0000-0001-9038-9950]{Schuyler D. Van Dyk}
\affil{Caltech/Spitzer Science Center, Caltech/IPAC, Mailcode 100-22, Pasadena, CA 91125, USA}

\author[0000-0001-5510-2424]{Nathan Smith}
\affil{Steward Observatory, University of Arizona, 933 North Cherry Avenue, Tucson, AZ 85721-0065, USA}

\author[0000-0001-7351-2531]{Jack M. M. Neustadt}
\affil{Department of Astronomy, The Ohio State University, 140 West 18th Avenue, Columbus, OH 43210, USA}

\author[0000-0003-4102-380X]{David J. Sand}
\affil{Steward Observatory, University of Arizona, 933 North Cherry Avenue, Tucson, AZ 85721-0065, USA}

\author{K. Kreckel}
\affil{Astronomisches Rechen-Institut, Zentrum f{\"u}r Astronomie der Universit{\"a}t Heidelberg, M{\"o}nchhofstra{\ss}e 12-14, 69120 Heidelberg, Germany}

\author{C. S. Kochanek}
\affil{Department of Astronomy, The Ohio State University, 140 West 18th Avenue, Columbus, OH 43210, USA}

\author[0000-0001-8818-0795]{S.~Valenti}
\affiliation{Department of Physics, University of California, 1 Shields Avenue, Davis, CA 95616-5270, USA}

\author[0000-0002-1468-9668]{Jay Strader}
\affil{Department of Physics and Astronomy, Michigan State University, East Lansing, MI 48824, USA}

\author{M. C. Bersten}
\affiliation{Instituto de Astrof\'isica de La Plata (IALP), CONICET, Paseo del bosque S/N, 1900, Argentina}
\affiliation{Facultad de Ciencias Astron\'omicas y Geof\'isicas, Universidad Nacional de La Plata, Paseo del Bosque, La Plata, Argentina}
\affiliation{Kavli Institute for the Physics and Mathematics of the Universe (WPI), The University of Tokyo, 5-1-5 Kashiwanoha, Kashiwa, Chiba, Japan}

\author{Guillermo A. Blanc}
\affil{The Observatories of the Carnegie Institution for Science, 813 Santa Barbara Street, Pasadena, CA 91101, USA}
\affil{Departamento de Astronom\'{i}a, Universidad de Chile, Casilla 36-D, Santiago, Chile}

\author[0000-0002-4924-444X]{K. Azalee Bostroem}
\affiliation{Department of Physics, University of California, 1 Shields Avenue, Davis, CA 95616-5270, USA}

\author{Thomas G. Brink}
\affiliation{Department of Astronomy, University of California, Berkeley, CA 94720-3411, USA.}

\author{Eric Emsellem}
\affiliation{European Southern Observatory, Karl-Schwarzschild-Stra{\ss}e 2, D-85748 Garching bei M\"{u}nchen, Germany}
\affiliation{Univ.\ Lyon, ENS de Lyon, CNRS, Centre de Recherche Astrophysique de Lyon UMR5574, F-69230 Saint-Genis-Laval France}

\author{Alexei V. Filippenko}
\affiliation{Department of Astronomy, University of California, Berkeley, CA 94720-3411, USA.}
\affiliation{Miller Institute for Basic Research in Science,
University of California, Berkeley, CA  94720  USA}

\author{Gast\'on Folatelli}
\affiliation{Instituto de Astrof\'isica de La Plata (IALP), CONICET, Paseo del bosque S/N, 1900, Argentina}
\affiliation{Facultad de Ciencias Astron\'omicas y Geof\'isicas, Universidad Nacional de La Plata, Paseo del Bosque, La Plata, Argentina}
\affiliation{Kavli Institute for the Physics and Mathematics of the Universe (WPI), The University of Tokyo, 5-1-5 Kashiwanoha, Kashiwa, Chiba, Japan}

\author[0000-0002-5619-4938]{Mansi M. Kasliwal}
\affil{Division of Physics, Mathematics, and Astronomy, California Institute of Technology, Pasadena, CA 91125, USA}

\author[0000-0002-8532-9395]{Frank J. Masci}
\affil{Caltech/Spitzer Science Center, Caltech/IPAC, Mailcode 100-22, Pasadena, CA 91125, USA}

\author{Rebecca McElroy}
\affiliation{Sydney Institute for Astronomy, School of Physics, A28, The University of Sydney, NSW, 2006, Australia}

\author{Dan Milisavljevic}
\affil{Department of Physics and Astronomy, Purdue University, 525 Northwestern Avenue, West Lafayette, IN 47907, USA}

\author{Francesco Santoro}
\affil{Max Planck Institut f\"{u}r Astronomie, K\"{o}nigstuhl 17, D-69117 Heidelberg, Germany
}

\author[0000-0003-4610-1117]{Tam\'as Szalai}
\affiliation{Department of Optics and Quantum Electronics, University of Szeged, H-6720 Szeged, D\'om t\'er 9., Hungary}
\affiliation{Konkoly Observatory, Research Centre for Astronomy and Earth Sciences, H-1121 Budapest, Konkoly Thege Mikl\'os \'ut 15-17,
Hungary}

\begin{abstract}

Extensive archival \textit{Hubble Space Telescope}, \textit{Spitzer Space Telescope}, and Large Binocular Telescope imaging of the recent intermediate-luminosity transient, AT~2019krl in M74, reveal a bright optical and mid-infrared progenitor star. While the optical peak of the event was missed, a peak was detected in the infrared with an absolute magnitude of $M_{4.5\,\mu {\rm m}} = -18.4$ mag, leading us to infer a visual-wavelength peak absolute magnitude of $-$13.5 to $-$14.5.  The pre-discovery light curve indicated no outbursts over the previous 16\,yr. The colors, magnitudes, and inferred temperatures of the progenitor best match a 13--14 M$_{\sun}$ yellow or blue supergiant (BSG), if only foreground extinction is taken into account, or a hotter and more massive star, if any additional local extinction is included.  A pre-eruption spectrum of the star reveals strong H$\alpha$ and [N~{\sc ii}] emission with wings extending to $\pm 2000$\,km\,s$^{-1}$. The post-eruption spectrum is fairly flat and featureless with only H$\alpha$, \ion{Na}{1}~D, [\ion{Ca}{2}], and the \ion{Ca}{2} triplet in emission. As in many previous intermediate-luminosity transients, AT~2019krl shows remarkable observational similarities to luminous blue variable (LBV) giant eruptions, SN~2008S-like events, and massive-star mergers. However, the information about the pre-eruption star favors either a relatively unobscured BSG or a more extinguished LBV with $M > 20$\,M$_{\sun}$ likely viewed pole-on.

\end{abstract}


\section{Introduction} \label{sec:intro}

Existing in the magnitude space between traditional supernovae (SNe) and classical novae lies a menagerie of explosive and eruptive transients with peak magnitudes in the range $-10 < M_{V} < -15$ mag and optical spectra dominated by narrow- or intermediate-width Balmer emission lines. These  ``SN imposters" \citep{2000PASP..112.1532V,2011MNRAS.415..773S,2012ApJ...758..142K,VanDyk+2012} may arise from a variety of progenitors and have been attributed to a number of potential physical mechanisms, including instabilities near the Eddington limit \citep{1994PASP..106.1025H,2006ApJ...645L..45S,2004ApJ...616..525O}, instabilities in nuclear burning in late post-main-sequence evolution \citep{2014ApJ...780...96S,2014ApJ...785...82S}, stellar mergers or common-envelope phases in binary star systems \citep{2013ApJ...764L...6S,2014MNRAS.443.1319K,SmithNGC4490,2018MNRAS.480.1466S}, or electron-capture supernovae \citep[ecSNe;][]{Botticella2008S,2012ApJ...758..142K,2016MNRAS.460.1645A}. 

\begin{figure*}
\begin{center}
\includegraphics[width=\linewidth]{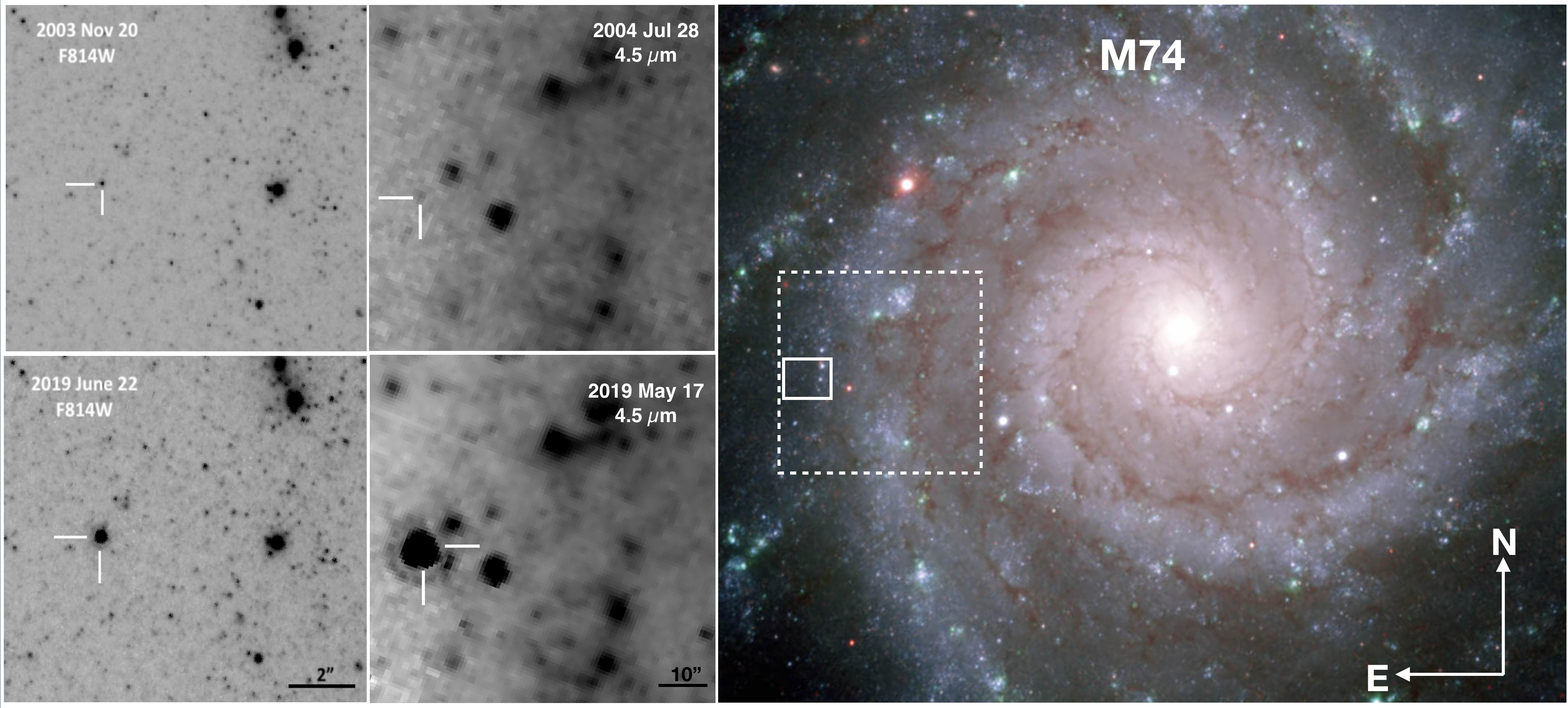}
\caption{Pre- and post-eruption images of AT~2019krl with \textit{HST} $F814W$ (two left panels) and {\it Spitzer}/IRAC 4.5\,$\mu$m (middle panels). AT~2019krl is indicated by the white tick marks, and the panels have the same orientation with north up, east to the left. The scale of the \textit{HST} (solid rectangle) and \textit{SST} (dashed rectangle) images is shown against the Gemini/GMOS color image on the right.}
\label{fig:finderchart}
\end{center}
\end{figure*}

Originally, the handful of known SN imposters were interpreted as giant eruptions of massive stars akin to $\eta$ Carinae's Great Eruption \citep{2000PASP..112.1532V,1989ApJ...342..908G,1995AJ....110.2261F,1999PASP..111.1124H,2001PASP..113..692S}. While giant eruptions are one type of outburst experienced by luminous blue variables (LBVs), they are phenomenologically different from the lower-amplitude, irregular, S-Doradus variations that are more commonly seen in LBVs \citep{2001A&A...366..508V,2005A&A...435..239C,2011MNRAS.415..773S,2017RSPTA.37560268S,2020Galax...8...20W,2020Galax...8...10D}.

Over the years, as more intermediate-luminosity transients have been discovered and a broader diversity was seen in their light curves, spectra, and possible progenitors, they were grouped into three broad classes of events: (1) giant eruptions of massive LBVs, (2) SN~2008S-like events (also known as intermediate-luminosity red transients, ILRTs, or intermediate-luminosity optical transients, ILOTs) that have been proposed as eruptions of heavily dust enshrouded blue supergiants or explosions of super-asymptotic-giant-branch (AGB) stars as ecSNe, and (3) luminous red novae (LRNe), which have usually been interpreted as binary mergers or common-envelope (CE) ejections in low- or intermediate-mass stars. All of these involve large amounts of episodic mass loss, and many of them share observed properties that blur the distinction between categories. For example, LBVs can experience super-Eddington eruptions which are accompanied by large amounts of mass loss \citep{2006ApJ...645L..45S,2004ApJ...616..525O}, but some LBV eruptions might also be the result of stellar mergers \citep{SmithNGC4490,2018MNRAS.480.1466S,Pastorello2019}. The most well known example of the phenomenon was the Great Eruption of $\eta$ Car \citep{2012ASSL..384..145S,2018MNRAS.480.1466S}. The SN~2008S-like events are characterized by a highly obscured dusty progenitor, and strong [\ion{Ca}{2}] and \ion{Ca}{2} near-infrared (NIR) triplet emission lines in their spectra \citep{Prieto2008S,2009ApJ...705.1425P,2009ApJ...705.1364T}, but some LBVs including $\eta$ Car exhibit all these properties as well \citep{2011MNRAS.415..773S,SmithNGC4490,2018MNRAS.480.1466S}. The SN 2008S-like transients have been interpreted as arising either from a terminal low-luminosity SN event \citep{Botticella2008S,2012ApJ...758..142K,2016MNRAS.460.1645A} or from massive-star outbursts in a dusty cocoon \citep{2009ApJ...699.1850B,2009ApJ...695L.154B,2009ApJ...697L..49S,2011ApJ...743..118H}.

With the discovery that the outburst of V1309 Sco was due to the merger of an inspiraling binary system of 1--2 M$_\sun$ \citep{2010A&A...516A.108M,2011A&A...528A.114T,Pejcha2014}, links could be made between red novae and merger events \citep{2011A&A...528A.114T}, including the more massive (3--10 M$_\sun$) proposed mergers V838 Mon \citep{2003Natur.422..405B,2008AJ....135..605S} and M31-LRN-2015 \citep{2015ATel.7173....1D,2017ApJ...835..282M,2020MNRAS.496.5503B}. The spectra of these events change dramatically with time, starting with narrow Balmer emission lines on top of a rather featureless blue continuum, and evolving to a cool, dusty, molecular-band-dominated spectrum.  Other well-known mergers of even more massive stars include NGC~4490-OT at $\sim 30$~M$_\sun$ \citep{SmithNGC4490,Pastorello2019}  and the similar transient AT~2017jfs \citep{2019A&A...625L...8P}, M101-2015OT1 at $\sim$18~M$_\sun$ \citep{2017ApJ...834..107B,2016AstBu..71...82G}, and SNHunt248 with a mass possibly as large as 60~M$_\sun$ \citep{2018MNRAS.473.3765M}. The light curves of these objects show prominent double or multiple peaks, with more massive progenitors linked with brighter peak magnitudes and a longer duration between peaks \citep{2014MNRAS.443.1319K,SmithNGC4490,Pastorello2019}.

Some intermediate-luminosity transients cannot be strictly classified into one of the three groups discussed above.  For instance, UGC~2773-OT exhibited [\ion{Ca}{2}] and \ion{Ca}{2} emission in its spectra, similar to the SN~2008S-like events, but appears to have had a luminous, blue progenitor and a slow rise to peak luminosity and a decade-long eruption akin to the Great Eruption of $\eta$ Car \citep{2010AJ....139.1451S,2016MNRAS.455.3546S,2011ApJ...732...32F}.  Moreover, $\eta$ Car --- the quintessential LBV giant eruption --- showed prominent [\ion{Ca}{2}] emission and molecular absorption in light-echo spectra \citep{2014ApJ...787L...8P,2018MNRAS.480.1466S}, plus prodigious dust formation and other features that are also attributed to ILRTs. Similarly, the optical spectra of SN~2002bu evolved from the appearance of an LBV to that of a SN~2008S type, and observations over a decade after the outburst are still inconclusive about whether the event was terminal \citep{2012ApJ...760...20S}.

Here we present another case of an intermediate-luminosity transient that shows outburst characteristics belonging to LBV, SN~2008S-like events, and massive star mergers. In this case, however, a luminous blue progenitor is clearly detected in pre-eruption data. AT~2019krl (ZTF19abehwhj) was discovered on 2019 July 07 \citep{2019TNSTR1165....1H} by the Zwicky Transient Facility \citep[ZTF;][]{2019PASP..131a8002B} in the nearby spiral galaxy M74 (NGC~628).
It was later classified as either a Type IIn supernova  or an LBV in outburst, based on an optical spectrum taken on 2019 July 8.4 that showed strong, complex H$\alpha$ emission with a narrow (130\,km\,s$^{-1}$) and an intermediate (2000\,km\,s$^{-1}$) width component \citep{2019ATel12913....1A}. M74 has been host to the well-studied SNe 2002ap, 2003gd, and 2013ej which have resulted in a rich dataset of archival imaging in the optical and infrared. From ground-based imaging using 20 reference Gaia DR2 \citep{GAIA2018} stars we obtained an absolute position of AT~2019krl of $\alpha$(J2000) $=$ 01$^h$36$^m$49$^s$.633, $\delta$(J2000) $=15\degr 46\arcmin 46\farcs 32$. 
A subsequent search of the \textit{Spitzer} Heritage Archive found that the object was detected in archival \textit{Spitzer Space Telescope} \citep[{\it SST};][]{werner04,gehrz07} images and appeared as a bright source in the last observational epoch on 2019 May 17, approximately two months prior to the optical discovery \citep{2019ATel12934....1S}. Adopting a distance modulus to M74 of $\mu$ = 29.95 $\pm$ 0.03 (stat.) $\pm$0.07 (syst.) mag  \citep[$d = 9.77\pm0.17\pm0.32$ Mpc;][which is consistent with the distance determined by \citealt{Kreckel2017} using the planetary nebula luminosity function]{2017AJ....154...51M}, the absolute magnitude of AT~2019krl in the brightest epoch from \textit{Spitzer} was $M_{4.5\,\mu {\rm m}}$ = $-$18.4. The combination of bright mid-infrared (MIR) emission, low optical brightness, and narrow Balmer emission suggested that AT~2019krl was likely one of the intermediate-luminosity transients discussed above. 

We outline the observations and data reduction in Section \ref{sec:observations}, and discuss the light curve and spectroscopic evolution of the progenitor and event in Section \ref{sec:analysis}. Section \ref{Results} discusses the constraints on the progenitor and explosion from the data, and Section \ref{discussion} compares these with other intermediate-luminosity transient types. We end with concluding remarks in Section \ref{sec:conclusions}.

\begin{figure*}
\begin{center}
\includegraphics[width=.48\linewidth]{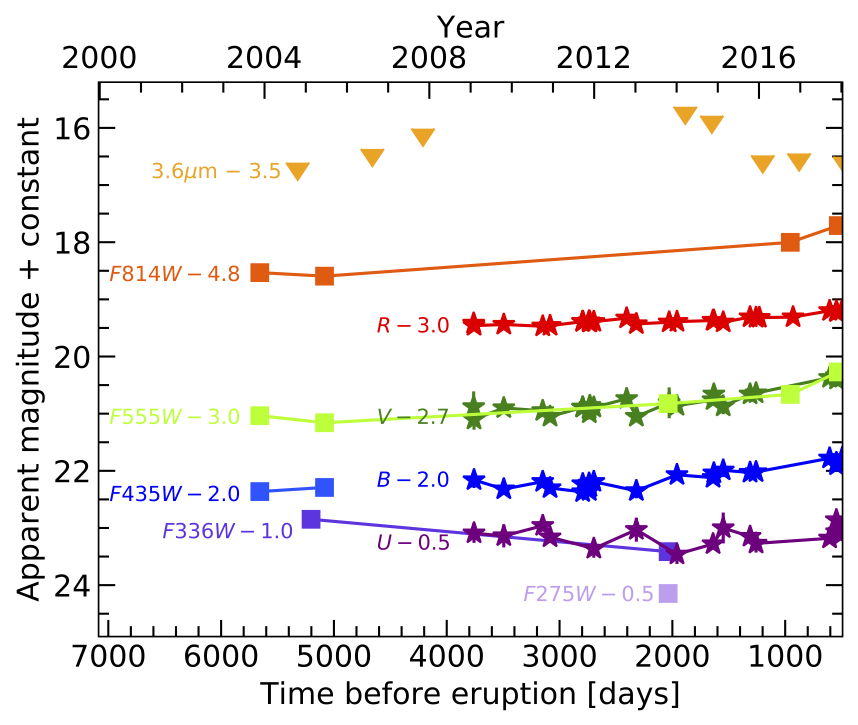}
\includegraphics[width=.45\linewidth]{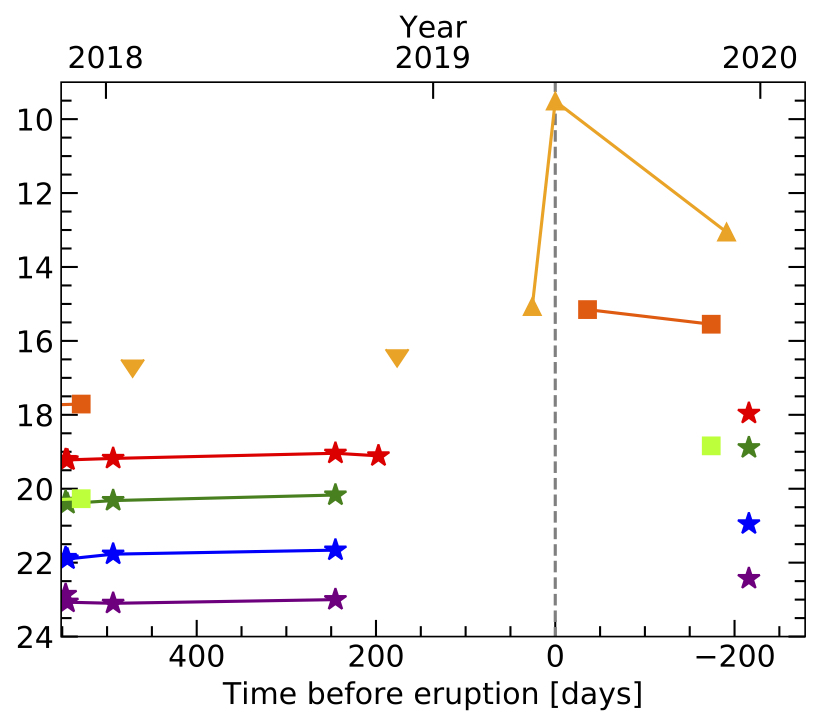}
\caption{Optical and infrared light curves of AT~2019krl. \textit{UBVR} data are from the LBT and other data are from \textit{HST} and \textit{Spitzer}. The light curves have been shifted by the constants indicated for ease of viewing. The left panel shows the  light curve of the progenitor, while the right panel focuses on the eruption. The date of our brightest \textit{Spitzer} epoch is indicated by a vertical dashed line, and upper limits from the \textit{Spitzer} measurements, stacked in one-year bins, are indicated by downward pointing triangles. The upper limits are similar in the 3.6 $\mu$m and 4.5 $\mu$m bands. The data are presented in Tables \ref{tab:measurements}, \ref{tab:LBTphot}, and \ref{tab:Spitzerphot}.}
\label{fig:LC}
\end{center}
\end{figure*}

\section{Observations} \label{sec:observations}
\subsection{\textit{HST} Photometry}
The site of the transient has been imaged many times before with \textit{HST}.  In addition, \textit{HST}/ACS $F814W$ observations of M74 obtained  on 2019 June 22 (PI: D. Sand) serendipitously imaged AT~2019krl two weeks before the discovery report was issued \citep{2019TNSTR1165....1H}. Using this post-outburst observation,  we could easily isolate the progenitor star in  pre-outburst archival \textit{HST} images. 

Pre-transient ACS/WFC data were obtained from programs GO-9796 (PI: J.~Miller; 2003 November 20), GO-10402 (PI: R.~Chandar; 2005 June 16), and GO-15645  (PI: D.~Sand; 2019 June 22). Several epochs of WFC3/UVIS are available, including from programs GO-13364 (PI:  D.~Calzetti; 2013 October 17), GO-13773 (PI: R.~Chandar; 2014 October 14), GO-14668 (PI: A.~Filippenko; 2016 October 04), and GO-15166 (PI: A.~Filippenko; 2017 December 04). Additionally, another  post-explosion epoch was taken with WFC3/UVIS on 2019 November 07 (GO-15151; PI: S. Van Dyk). One epoch of WFPC2/WF3 data was also obtained from GO-10402 (PI: R.~Chandar; 2005 February 16). The data were all obtained from the Mikulski Archive for Space Telescopes (MAST\footnote{https://archive.stsci.edu/}) with standard pipeline calibrations applied. See Table \ref{tab:measurements}. In Figure \ref{fig:finderchart} we show the transient location in a pre-eruption image from 2003, and one post-eruption image from 2019. We analyzed these data with DOLPHOT\footnote{http://americano.dolphinsim.com/dolphot/} \citep{2000PASP..112.1383D,dolphin2016},  after using  AstroDrizzle \citep{Hack2012} to produce drizzled image mosaics and to flag cosmic-ray hits in the individual frames. We used the recommended parameters for DOLPHOT and adopted values for the parameters \texttt{FitSky}=3 and \texttt{RAper}=8 for the photometry. We present the \textit{HST} photometry on the Vega scale in Table \ref{tab:measurements}.

\begin{table}
\begin{center}
\begin{minipage}{\linewidth}
      \caption{\textit{HST} observations}
\begin{tabular}{@{}lccc}\hline\hline
Date &Filter   & Instrument  & VegaMAG\tablenotemark{a}   \\ 
 \hline       
2003-11-20 & $F435W$ & ACS/WFC  &24.363$\pm$0.012 \\
 & $F555W$ & ACS/WFC  &24.035$\pm$0.016 \\
 & $F814W$ & ACS/WFC  &23.332$\pm$0.015 \\
 2005-02-16 & $F336W$ & WFPC2/WF3 &23.848$\pm$0.121 \\
2005-06-16 & $F435W$ & ACS/WFC  &24.291$\pm$0.042 \\
 & $F555W$ & ACS/WFC &24.159$\pm$0.045 \\
 & $F814W$ & ACS/WFC &23.394$\pm$0.026 \\
2013-10-17 & $F275W$ &WFC3/UVIS  & 24.646$\pm$0.115 \\
& $F336W$ &WFC3/UVIS  & 24.414$\pm$0.094 \\
& $F555W$ &WFC3/UVIS  & 23.824$\pm$0.020 \\
2014-10-14 & $F547M$ &WFC3/UVIS  & 23.713$\pm$0.044 \\
& $F657N$ &WFC3/UVIS &  21.089$\pm$0.022 \\
2016-10-04 & $F555W$ &WFC3/UVIS & 23.663$\pm$0.021 \\
& $F814W$ &WFC3/UVIS &  22.802$\pm$0.024 \\
2017-12-04 & $F555W$ &WFC3/UVIS &  23.270$\pm$0.018 \\
& $F814W$ &WFC3/UVIS &  22.509$\pm$0.022 \\
2019-06-22 & $F814W$ & ACS/WFC & 19.953$\pm$0.003 \\
2019-11-07 & $F555W$ &WFC3/UVIS &  21.840$\pm$0.025 \\
& $F814W$ &WFC3/UVIS &  20.349$\pm$0.020 \\

\hline
\end{tabular}\label{tab:measurements}
\tablenotetext{a}{DOLPHOT magnitudes obtained from the \textit{HST} data.}
\end{minipage}
\end{center}
\end{table}

\subsection{LBT Photometry}
\begin{table*}
\centering
\caption{AT~2019krl LBT photometry\tablenotemark{a}}
\begin{tabular}{@{}lcccc}\hline\hline
MJD &$U$ & $B$  & $V$ &  $R$  \\
& mag&mag&mag&mag\\
 \hline  

54859 & -- & -- & 23.58 $\pm$ 0.27 & 22.42 $\pm$ 0.06\\
54862 & 23.59 $\pm$ 0.17 & 24.16 $\pm$ 0.14 & 23.80 $\pm$ 0.18 & 22.46 $\pm$ 0.01\\
55126 & 23.64 $\pm$ 0.20 & 24.32 $\pm$ 0.06 & 23.60 $\pm$ 0.07 & 22.44 $\pm$ 0.01\\
55471 & 23.46 $\pm$ 0.08 & 24.19 $\pm$ 0.04 & 23.65 $\pm$ 0.05 & 22.47 $\pm$ 0.01 \\
55536 & 23.66 $\pm$ 0.05 & 24.30 $\pm$ 0.04 & 23.75 $\pm$ 0.04 & 22.46 $\pm$ 0.01 \\
55825 & -- &  24.37 $\pm$ 0.12 & 23.59 $\pm$ 0.06 & 22.39 $\pm$ 0.02\\
55826 & -- &  24.23 $\pm$ 0.04 & 23.59 $\pm$ 0.03 & 22.39 $\pm$ 0.02\\
55882 & -- & 24.22 $\pm$ 0.07 & 23.62 $\pm$ 0.06 & 22.38 $\pm$ 0.02\\
55884 & -- &  24.35 $\pm$ 0.08 & 23.69 $\pm$ 0.11 & 22.40 $\pm$ 0.01\\
55889 & -- &  24.28 $\pm$ 0.07 & 23.56 $\pm$ 0.05 & 22.40 $\pm$ 0.02\\
55924 & 23.86 $\pm$ 0.16 & 24.19 $\pm$ 0.06 & 23.59 $\pm$ 0.03 & 22.39 $\pm$0.01\\
56215 & -- & --& 23.44 $\pm$ 0.13 & 22.33 $\pm$ 0.07\\
56301 & 23.53 $\pm$ 0.21 & 24.35 $\pm$ 0.13 & 23.75 $\pm$ 0.07 & 22.43 $\pm$ 0.02\\
56592 & -- & --  & 23.51 $\pm$ 0.27 & 22.40 $\pm$ 0.03\\
56661 & 23.96 $\pm$ 0.17 & 24.07 $\pm$ 0.04 & 23.56 $\pm$ 0.05 & 22.39 $\pm$ 0.01\\
56981 & 23.78 $\pm$ 0.09 & 24.12 $\pm$ 0.03 & 23.46$\pm$ 0.03 & 22.37 $\pm$ 0.01\\
56988 & -- & 24.04 $\pm$ 0.08 & 23.36 $\pm$ 0.07 & 22.37 $\pm$ 0.01\\
57071 & 23.50 $\pm$ 0.27 & 23.99 $\pm$ 0.12 & 23.57 $\pm$ 0.14 & 22.40 $\pm$ 0.03\\
57309 & 23.65 $\pm$ 0.07 & 24.03 $\pm$ 0.06 & 23.36 $\pm$ 0.03 & 22.31 $\pm$ 0.01\\
57362 & 23.77 $\pm$ 0.16 & 24.02 $\pm$ 0.07 & 23.34 $\pm$ 0.06 & 22.32 $\pm$ 0.02\\
57391 & -- & -- & -- & 22.32 $\pm$ 0.03\\
57690 & -- & -- & -- & 22.31 $\pm$ 0.02\\
58014 & 23.68 $\pm$ 0.05 & 23.78 $\pm$ 0.04 & 23.07 $\pm$ 0.03 & 22.20 $\pm$ 0.01\\
58074 & 23.36 $\pm$ 0.12 & 23.86 $\pm$ 0.04 & 23.04 $\pm$ 0.02 & 22.21 $\pm$ 0.01\\
58076 & 23.57 $\pm$ 0.16 & 23.90 $\pm$ 0.05 & 23.10 $\pm$ 0.08 & 22.22 $\pm$ 0.02\\
58127 & 23.60 $\pm$ 0.11 & 23.77 $\pm$ 0.04 & 23.02 $\pm$ 0.03 & 22.18 $\pm$ 0.01\\
58375 & 23.50 $\pm$ 0.06 & 23.66 $\pm$ 0.02 & 22.87 $\pm$ 0.03 & 22.04 $\pm$ 0.01\\
58423 & -- & -- & -- & 22.11 $\pm$ 0.03\\
58837 & 22.93 $\pm$ 0.10 & 22.95 $\pm$ 0.02 & 21.59 $\pm$ 0.01 & 20.96 $\pm$ 0.01\\

\hline
\end{tabular}\label{tab:LBTphot}
\tablenotetext{a}{Magnitudes are in the Vega system.}
\centering
\end{table*}

Observations of M74, including the position of AT~2019krl, were obtained as part of the Large Binocular Telescope (LBT) Search for Failed Supernovae \citep{kochanek08}.  As part of this survey,  \textit{UBVR} images of M74 were obtained between 2008 and 2019 using the Large Binocular Cameras \citep[LBC;][]{giallongo08} on the LBT.  The data reduction and image processing are described by \citet{gerke15} and \citet{adams17b}.  In summary, the best images are combined to make a reference image, and the individual epochs are analyzed using the ISIS image-subtraction package \citep{alard98,alard00}.  The difference imaging provides a light curve of the variable flux that is unaffected by crowding. The mean flux of the source in the reference image is subject to the effects of crowding and is less well-determined.

The data are calibrated using stars in the Sloan Digital Sky Survey \citep[SDSS;][]{ahn12}  and transformed to \textit{UBV$R_C$} Vega magnitudes using the conversions reported by \citet{jordi06}. These calibrations are accurate to 0.1~mag or better.  The uncertainties in the transient light curve are estimated using the variance of light curves extracted from nearby source-free regions, as these empirical uncertainties will include any systematic contributions to the uncertainties beyond simple Poisson errors. The LBT photometry is listed in Table \ref{tab:LBTphot}. 

\subsection{\textit{Spitzer} Photometry}
\begin{table*}
\begin{center}
\caption{AT~2019krl \textit{Spitzer} Photometry}
\begin{tabular}{@{}lrrrrrrrr}\hline\hline
MJD & [3.6] Diff.\ Flux & Error & [4.5] Diff.\ Flux & Error & [3.6] & Error & [4.5] & Error \\ 
    & ($\mu$Jy) & ($\mu$Jy) & ($\mu$Jy) & ($\mu$Jy) & (mag)     & (mag)     & (mag)     & (mag) \\
 \hline  
53211.82 & $    1.2$ & $    1.9$ & $    1.1$ & $    0.9$ & $>  19.3$ & \nodata   & $>  19.4$ & \nodata   \\
53385.98 & $   -1.3$ & $    1.8$ & $   -2.8$ & $    2.1$ & $>  19.3$ & \nodata   & $>  18.5$ & \nodata   \\
53960.85 & $    0.7$ & $    1.7$ & $   -0.4$ & $    1.1$ & $>  19.4$ & \nodata   & $>  19.1$ & \nodata   \\
54328.12 & $    0.1$ & $    3.2$ & $   -4.5$ & $    2.8$ & $>  18.6$ & \nodata   & $>  18.2$ & \nodata   \\
54491.19 & $   -2.5$ & $    3.1$ & $   -3.8$ & $    2.5$ & $>  18.7$ & \nodata   & $>  18.3$ & \nodata   \\
56734.98 & $    0.6$ & $    3.2$ & $   -1.3$ & $   12.6$ & $>  18.6$ & \nodata   & $>  16.7$ & \nodata   \\
56936.57 & \nodata   & \nodata   & $    3.4$ & $    1.5$ & \nodata   & \nodata   & $>  18.8$ & \nodata   \\
56970.14 & $   -0.4$ & $    2.8$ & $    0.0$ & $    1.3$ & $>  18.8$ & \nodata   & $>  19.0$ & \nodata   \\
57312.98 & $   -1.1$ & $    2.8$ & \nodata   & \nodata   & $>  18.8$ & \nodata   & \nodata   & \nodata   \\
57320.53 & \nodata   & \nodata   & $    1.0$ & $    1.8$ & \nodata   & \nodata   & $>  18.6$ & \nodata   \\
57334.24 & $    0.9$ & $    3.3$ & $    0.0$ & $    1.7$ & $>  18.6$ & \nodata   & $>  18.7$ & \nodata   \\
57474.90 & $    1.3$ & $    2.1$ & \nodata   & \nodata   & $>  19.1$ & \nodata   & \nodata   & \nodata   \\
57482.44 & $    4.7$ & $   21.8$ & $    2.1$ & $    2.9$ & $>  16.6$ & \nodata   & $>  18.2$ & \nodata   \\
57503.57 & $    0.7$ & $    4.3$ & $   -1.7$ & $   17.1$ & $>  18.3$ & \nodata   & $>  16.3$ & \nodata   \\
57680.70 & $    4.2$ & $    2.4$ & $    3.6$ & $    2.8$ & $>  19.0$ & \nodata   & $>  18.2$ & \nodata   \\
57695.05 & $   -0.3$ & $    2.3$ & $   -0.3$ & $    1.7$ & $>  19.0$ & \nodata   & $>  18.7$ & \nodata   \\
57855.38 & $    1.3$ & $    3.3$ & \nodata   & \nodata   & $>  18.6$ & \nodata   & \nodata   & \nodata   \\
58054.61 & $    2.9$ & $    1.7$ & $    2.5$ & $    0.9$ & $>  19.4$ & \nodata   & $>  19.3$ & \nodata   \\
58242.87 & $    5.8$ & $    3.2$ & $   10.4$ & $   20.0$ & $>  18.7$ & \nodata   & $>  16.2$ & \nodata   \\
58427.87 & $    2.8$ & $    2.1$ & $    8.2$ & $    1.2$ & $>  19.1$ & \nodata   & $  18.35$ & $   0.16$ \\
58459.75 & $    4.9$ & $    4.0$ & $   10.5$ & $    2.7$ & $>  18.4$ & \nodata   & $  18.08$ & $   0.28$ \\
58594.60 & $   16.6$ & $    1.8$ & $   29.8$ & $    2.1$ & $  18.07$ & $   0.12$ & $  16.95$ & $   0.08$ \\
58620.24 & $ 2779.3$ & $   15.1$ & $ 4121.7$ & $   25.4$ & $  12.51$ & $   0.01$ & $  11.60$ & $   0.01$ \\
58811.34 & $  106.1$ & $    3.1$ & $  216.4$ & $    1.3$ & $  16.06$ & $   0.03$ & $  14.80$ & $   0.01$ \\

\hline
\end{tabular}\label{tab:Spitzerphot}

\end{center}
\end{table*}

There have been many observations of M74 in the 3.6 and 4.5~$\mu$m imaging channels ([3.6] and [4.5]) of the Infrared Array Camera (IRAC; \citealp{fazio04}) on-board \textit{Spitzer}  since 2004 as part of several observing programs (PID 159, PI: R.\ Kennicutt; PID 3248, PI: W.\ P.\ Meikle; PID 30494, PI: B.\ Sugerman; PID 40010, PI: M.\ Meixner), including extensive coverage since 2014 by the SPitzer InfraRed Intensive Transients Survey (SPIRITS; PIDs 10136, 11063, 13053, 14089; PI: M.\ Kasliwal) through the end of 2019. Pre-discovery photometry was presented by \citet{2019ATel12934....1S} up until the infrared (IR) peak of the transient on 2019 May 17, including the upper limits of the nondetections at 5.8\,$\mu$m and 8.0\,$\mu$m of 5\,$\mu$Jy and 15\,$\mu$Jy, respectively.

As part of SPIRITS, the post-basic calibrated data (PBCD) level images were downloaded from the \textit{Spitzer} Heritage Archive\footnote{\url{https://sha.ipac.caltech.edu/applications/Spitzer/SHA/}} and \textit{Spitzer} Early Release Data Service\footnote{\url{http://ssc.spitzer.caltech.edu/warmmission/sus/mlist/archive/2015/msg007.txt}} and processed through an automated image-subtraction pipeline (for details, see \citealp{kasliwal17}). For reference images, we used the images taken on 2004 July 28 for the \textit{Spitzer} Infrared Nearby Galaxies Survey (SINGS; \citealp{kennicutt03}).  We performed aperture photometry on the difference images using a 4 mosaicked-pixel ($2\farcs4$) aperture and background annulus from 4--12 pixels ($2\farcs4$--$7\farcs2$). The extracted flux is multiplied by the aperture corrections of 1.215 for [3.6] and 1.233 for [4.5] as described in the IRAC Instrument Handbook\footnote{\url{http://irsa.ipac.caltech.edu/data/SPITZER/docs/irac/iracinstrumenthandbook/}}. To estimate the photometric uncertainties, we performed photometry with the same parameters as above in a grid of apertures spanning a $32\arcsec$ box with $8\arcsec$ spacing centered at the location of the transient, excluding the central aperture. We adopt a robust estimate of the root-mean-square (rms) uncertainty in the distribution of flux measurements for the aperture grid ($0.5\times[85^{\mathrm{th}} - 16^{\mathrm{th}}$ percentile]) as representative of the 1$\sigma$ uncertainties in our photometry. 

In the 2004 reference images used for subtraction, a possible quiescent counterpart is visible at both [3.6] and [4.5]. Our aperture photometry gives low-significance measurements of $F_{\nu,[3.6]} = 4.6 \pm 4.5$ and $F_{\nu,[4.5]} = 7.1 \pm 3.6$\,$\mu$Jy, consistent with 2005 January 15 measurements by \citet{2019ATel12934....1S}. Given the limited spatial resolution of \textit{Spitzer}/IRAC and the complicated background emission, it is not possible to rule out that the emission at the site is due to confusion with nearby, unrelated sources. Thus, we infer 3$\sigma$ limits on the IR flux of the precursor in 2004 of $F_{\nu,[3.6]} < 14$ and $F_{\nu,[4.5]} < 11$\,$\mu$Jy. We adopt our difference imaging measurements throughout the rest of this work with the caveat that they may underestimate the true source flux. We convert our flux measurements to Vega-system magnitudes using the zero-magnitude fluxes presented for each IRAC channel in the IRAC Instrument Handbook and list our photometry in Table \ref{tab:Spitzerphot}.

\subsection{Spectroscopy}

Multiple long-slit optical spectra were taken of AT~2019krl with various telescopes/instruments between July and November 2019. These include one epoch with Binospec \citep{2019PASP..131g5004F} on the 6.5\,m MMT telescope, one epoch with the Kast double spectrograph (Miller $\&$ Stone 1993) mounted on the Shane 3\,m telescope at Lick observatory, one epoch with the Goodman spectrograph \citep{goodman} on the 4.1\,m SOAR telescope, one epoch taken with the DEep Imaging Multi-Object Spectrograph \citep[DEIMOS]{2003SPIE.4841.1657F}  on the 10\,m Keck-II telescope at Maunakea, and a final epoch with the Multi-Object Double Spectrographs  \citep[MODS]{2010SPIE.7735E..0AP} on the twin 8.4\,m LBT at Mount Graham International Observatory. These spectra were reduced using standard techniques, including bias subtraction, flat fielding, cosmic ray rejection, local sky subtraction, and extraction of one-dimensional spectra. The MMT data were reduced using the Binospec pipeline \citep{2019PASP..131g5005K}. Most observations had the slit aligned along the parallactic angle to minimize differential light losses \citep{1982PASP...94..715F}.  Flux calibration was done with standard-star observations taken on the same night at similar airmass.

 A pre-outburst spectrum is serendipitously available from observations using the Very Large Telescope/Multi Unit Spectroscopic Explorer (VLT/MUSE) spectrograph \citep{Bacon2010} as part of the PHANGS\footnote{Physics at High Angular resolution in Nearby GalaxieS; \url{http://www.phangs.org}}-MUSE survey (E.~Emsellem et al., in prep.). This optical integral field unit provides a $1\arcmin \times 1\arcmin$ field of view with 0\farcs2 pixels and a typical spectral resolution of $\sim 2.5$\,\AA\ over the nominal wavelength range, covering 4800--9300\,\AA.  Observations of M74 \citep{Kreckel2018, Kreckel2019} were taken on 2018 November 13 and targeted the source position in three rotations, alternating with two sky pointings, for a total on-source integration time of 50\,min.  Data reduction is carried out using a pipeline wrapping around the MUSE data reduction pipeline \citep{2020A&A...641A..28W} and developed by the PHANGS team.\footnote{\url{https://github.com/emsellem/pymusepipe}}
A log of the spectroscopic observations is given in Table \ref{tab:spectroscopy}.

\begin{figure*}
\includegraphics[width=\linewidth]{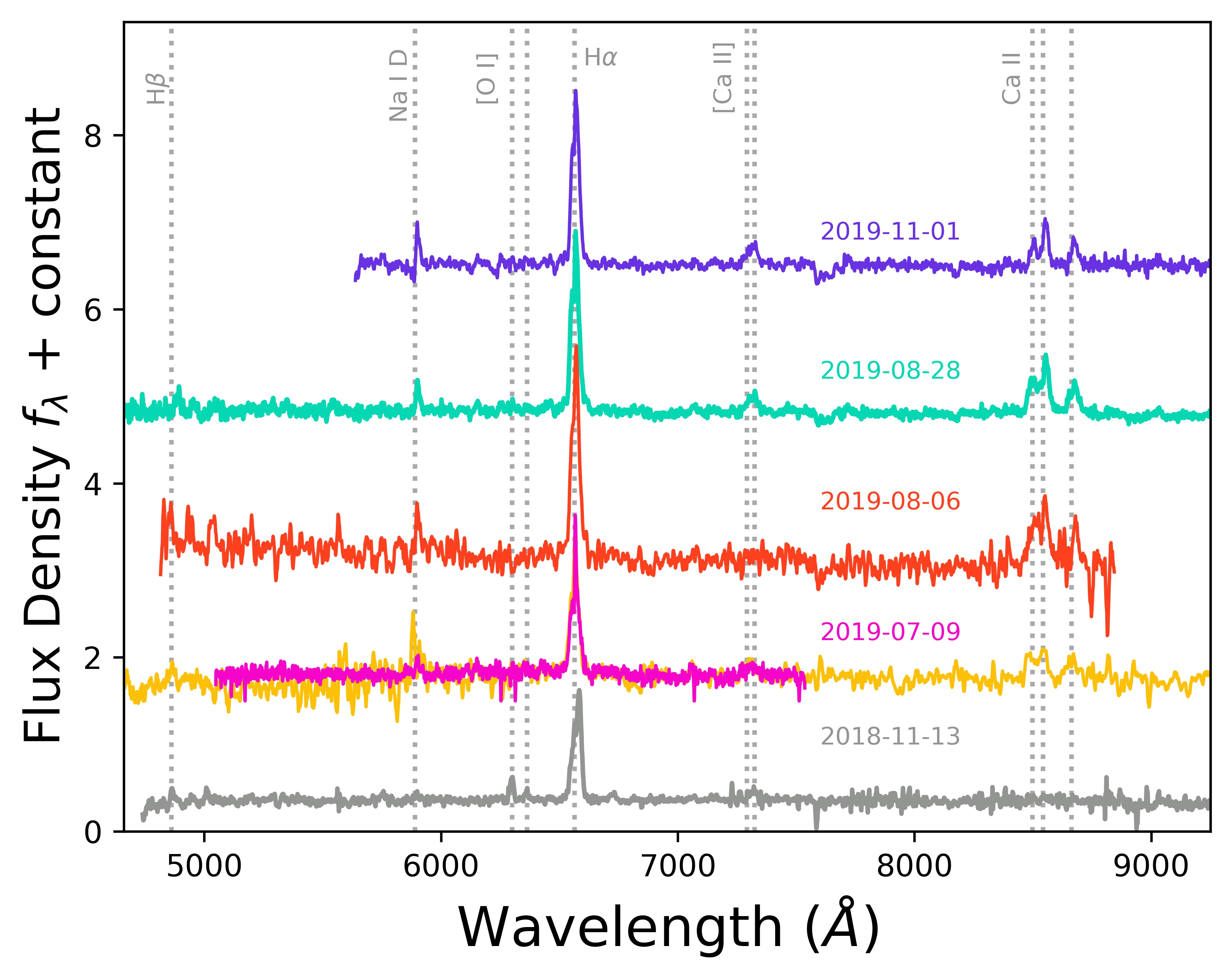}
\caption{Spectroscopic evolution of the progenitor (gray) and eruption of AT~2019krl.  A list of the spectroscopic observations is presented in Table \ref{tab:spectroscopy}.  The spectra have not been corrected for extinction, but have been smoothed to show prominent emission lines (which are marked with gray dotted lines) more clearly. }
\label{fig:spectra}
\end{figure*}

\begin{figure}
\includegraphics[width=\linewidth]{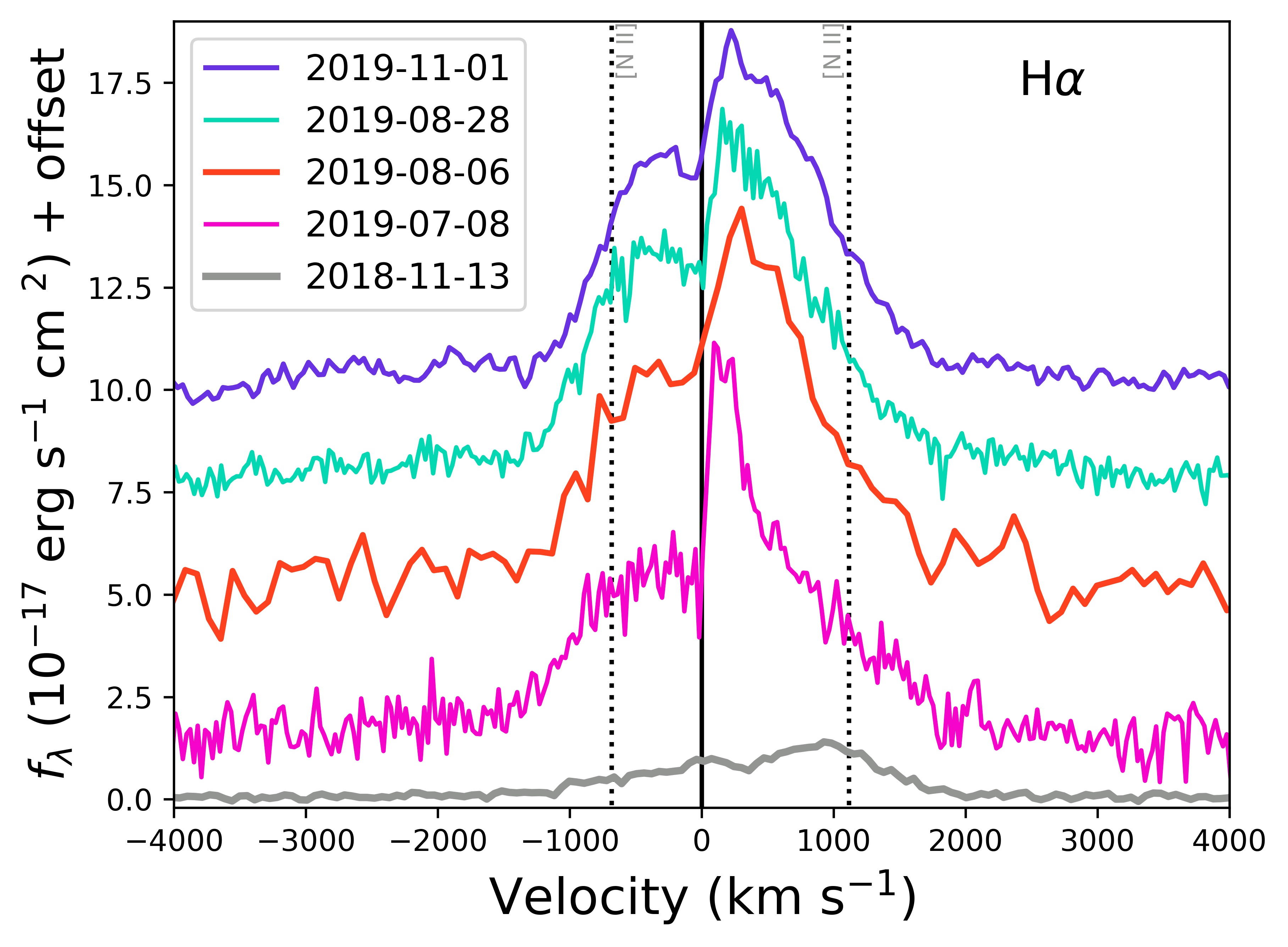}
\includegraphics[width=\linewidth]{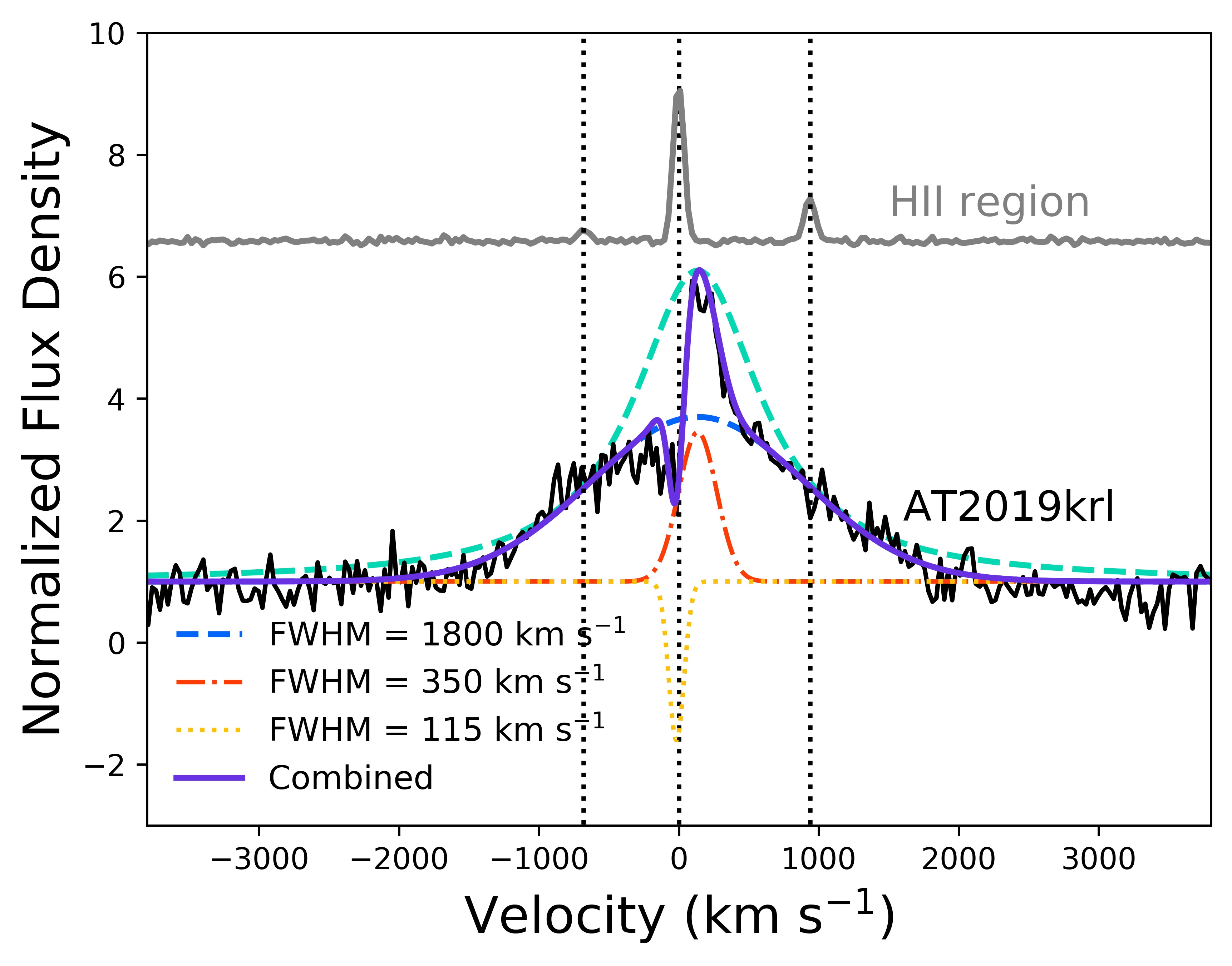}
\caption{Top: The evolution of the H$\alpha$ emission line, with zero velocity determined by the centroid of narrow H$\alpha$ emission from nearby H~{\sc ii} regions along the slit. The excess in the 2018 pre-eruption spectrum at $\sim 1000$\,km\,s$^{-1}$ is due to [\ion{N}{2}] $\lambda$6584 and is marked by a dashed line. [\ion{N}{2}] $\lambda$6548 is also marked near $-700$\,km\,s$^{-1}$. Bottom: The MMT/Binospec spectrum can be fit using 3 Gaussians: two in emission (narrow and broad), both centered at +135\,km\,s$^{-1}$ relative to nearby H~{\sc ii} regions, and one in absorption centered at $-20$\,km\,s$^{-1}$. The difference between emission and absorption velocities suggests an outflow of 155\,km\,s$^{-1}$. An additional Lorentzian (FWHM = $1100$\,km\,s$^{-1}$) is shown for comparison (teal dashed line).  The poor match to the line profile indicates that electron scattering does not dominate the production of the broad component.  The emission of a nearby \ion{H}{2} region is also shown to further illustrate the true redshift of the H$\alpha$ profile for AT~2019krl.}
\label{fig:halphacomp}
\end{figure}

\section{Analysis} \label{sec:analysis}
\subsection{Metallicity}
Using the adopted distance of 9.77 Mpc \citep{2017AJ....154...51M},  AT~2019krl is located roughly 5.4 kpc from the center of M74.  Assuming the oxygen abundance gradient in M74 is 12 + log[O/H]= $(8.834 \pm 0.069) + (-0.044 \pm 0.011) \times R$ dex\, kpc$^{-1}$ \citep{2015ApJ...806...16B} we derive  12~+~log[O/H] = $8.59 \pm 0.1$, a value consistent with the solar oxygen abundance of $8.69 \pm 0.05$ \citep{2009ARA&A..47..481A}. Therefore, we assume the metallicity at the location of AT~2019krl is approximately solar.

\subsection{Extinction}
The equivalent width (EW) of the \ion{Na}{1} D $\lambda\lambda$5889, 5896 absorption feature is often used following the prescription of \citet{2012MNRAS.426.1465P} to estimate the extinction of an extragalactic transient, although \citet{Phillips+2013} have cautioned against using this relation to obtain extinction estimates. Unfortunately, the \ion{Na}{1} D lines in AT~2019krl are seen only in emission (Figure \ref{fig:spectra}), likely from a contribution from the surrounding circumstellar medium (CSM).  For core-collapse SNe the observed color, for example, can be used to estimate the extinction, since the intrinsic colors of such SNe are relatively well defined \citep[e.g.,][although see  \citealt{dejaegar18}]{Drout2011,Stritzi2018}.  Since outbursts such as AT~2019krl are not well understood, this is also not a viable option.

However, we can instead attempt to constrain the reddening $E(B-V)$ of AT~2019krl from the nearby stellar population.  Using a technique similar to that outlined by \citet{2013ApJ...771...62K}, we use penalized pixel-fitting \citep{2004PASP..116..138C,2017MNRAS.466..798C} to determine the linear combination of \citet{2003MNRAS.344.1000B} simple stellar population templates that best fits an integrated 100\,pc wide annular integrated spectrum. This fit requires a third order multiplicative polynomial, which agrees well in shape with a \citet{2000ApJ...533..682C} attenuation law. From this comparison we obtain a value of $E(B-V)_{\rm total}$  = 0.12 mag, after 
including 
the Milky Way line-of-sight reddening toward M74 of $E(B-V)_{\rm MW}$  = 0.062 mag \citep{2011ApJ...737..103S}. This is only a lower limit, as circumstellar extinction around the transient may be much higher, but likely provides us with a reasonable estimate of total foreground extinction which we will use throughout the rest of the paper.

\subsection{Light Curve and Color Evolution}

The optical light curves, shifted for ease of viewing, are shown in Figure \ref{fig:LC}, with the photometry listed in Tables \ref{tab:measurements}, \ref{tab:LBTphot}, and \ref{tab:Spitzerphot}. The absolute magnitudes of the progenitor at the first epoch in 2003 are roughly $M_{F435W} = -$6.0 mag, $M_{F555W} = -$6.3 mag, and $M_{F814W} = -$6.8 mag corrected for $E(B-V)_{\rm total}$  = 0.12 mag.  In 2013 the progenitor is somewhat brighter with $M_{F275W} = -$5.9 mag, $M_{F336W} = -$6.1 mag, and $M_{F555W} = -$6.5 mag.  From our \textit{HST} photometry taken in 2017, we see that $M_{F555W} = -$7.0 mag, or almost a magnitude brighter than in 2003, and that between 2017 September and 2018 September it brightens by another 0.1--0.2 mag. The LBT data, which begin in 2009, show a fairly flat evolution up until late 2017, eliminating any other major eruptions in the decade previous.

The 3.6\,$\mu$m  magnitudes are also shown in Figure \ref{fig:LC}.  Only upper limits are obtained for the majority of the early evolution, but similar to the optical data, the 3.6\,$\mu$m and 4.5\,$\mu$m data do not seem to indicate any major outbursts between 2004 and 2018.  There is a noticeable increase from 2018 December to 2019 April as the 4.5\,$\mu$m luminosity increases from $-$11.9 to $-$13.0 mag. Finally, on 2019 May 17 we obtain our highest luminosities of $M_{3.6\,\mu {\rm m}}=-$17.5 mag, and $M_{4.5\,\mu {\rm m}}= -$18.4 mag.  From these \textit{Spitzer} data, we can constrain the peak of the outburst to be between 2019 April 21 and May 17.  The peak was not observed in the optical data owing to Sun constraints.

We only obtained a handful of observations after discovery.  The photometry from the ACS/$F814W$ image taken on 2019 June 22 reveals a luminosity of $M_{F814W} = -$10.2 mag, which then falls to $-$9.9 mag by 2019 November 07. Similarly,  $M_{3.6\,\mu {\rm m}}$ and $M_{4.5\,\mu {\rm m}}$ have dropped to $-$13.9 and $-$15.2 mag, respectively, by 2019  November, corresponding to a decrease of roughly 0.02 mag day$^{-1}$.

\begin{figure}
\begin{center}
\includegraphics[width=\linewidth]{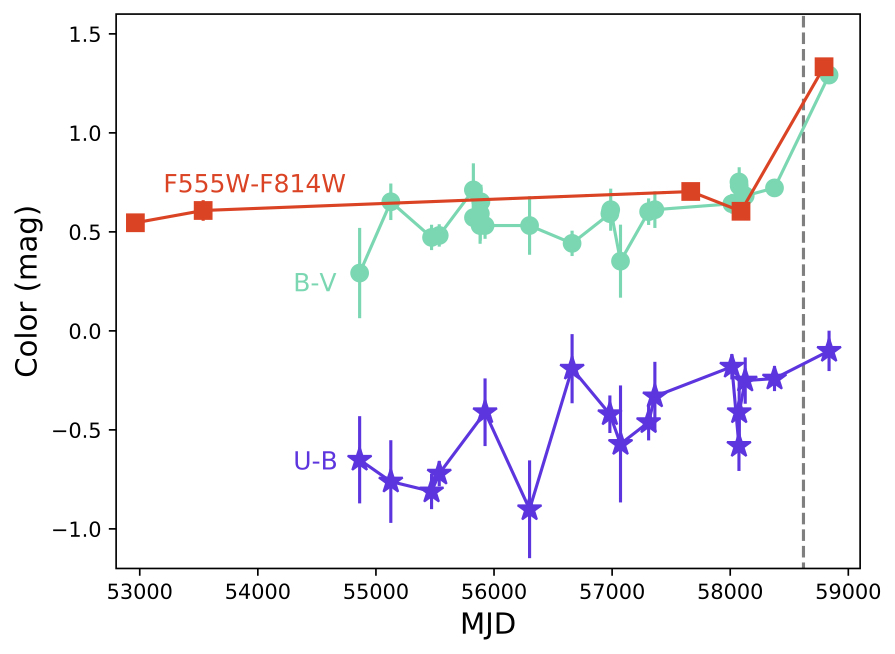}
\caption{Color evolution of the progenitor and outburst of AT~2019krl. As in other figures the date of our brightest \textit{Spitzer} epoch is indicated by a vertical dashed line and all data have been corrected for $E(B-V) =$ 0.12 mag.}
\label{fig:colorevol}
\end{center}
\end{figure}

As shown in Figure \ref{fig:colorevol}, the source steadily becomes redder, with a larger change in the color of the bluer bands.  The $U-B$ color evolves from roughly $-$0.6 to $-$0.1 mag, the $B-V$ from 0.4 to 0.8 mag, and the \textit{HST} $V-I$ color remains fairly flat at $\sim 0.7$ mag.  After the eruption the \textit{HST} $V-I$ and $B-V$ colors both jump to roughly 1.4 mag, while $U-B$ gets redder by only 0.1 mag. This indicates that the post-eruption object was much redder than the pre-eruption progenitor. We will discuss how the light curve and color evolution can be used to infer progenitor and explosion properties in Section \ref{Results} below.

\subsection{Spectroscopic Evolution}
The spectroscopic evolution of AT 2019krl, including a progenitor spectrum from $\sim 6$ months prior to eruption, are shown in Figure \ref{fig:spectra} and listed in Table \ref{tab:spectroscopy}. To confirm the rest velocities of the components which appear redshifted with respect to the zero velocity of the galaxy, we have also plotted the profile of a nearby \ion{H}{2} region in Figure \ref{fig:halphacomp}.  This exercise shows that there is a true velocity offset between the \ion{H}{2} region and the peak of H$\alpha$ emission, and that we are fully resolving the narrow H$\alpha$ component in AT~2019krl, which is much broader than the \ion{H}{2} region lines.

\begin{figure}
\includegraphics[width=\linewidth]{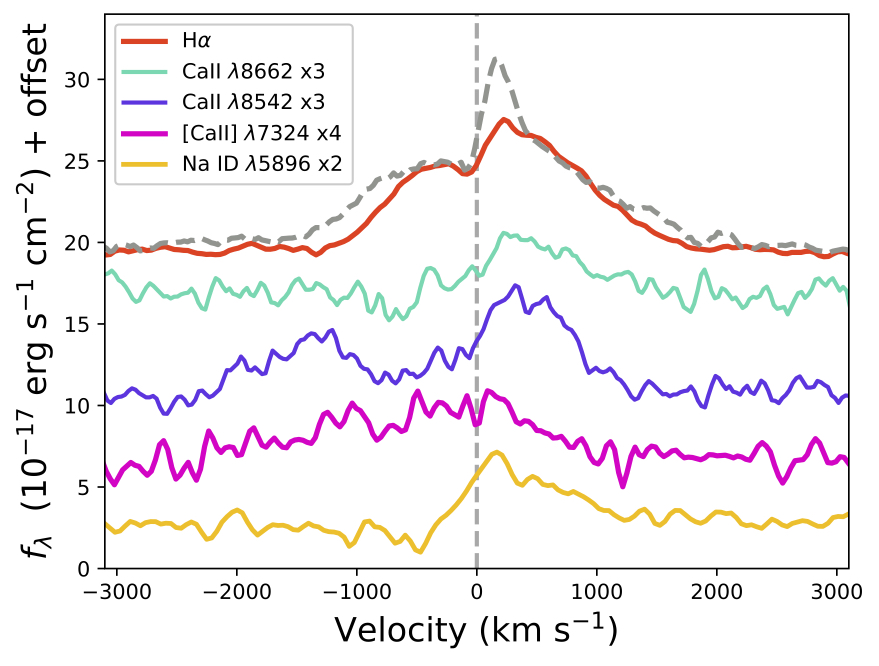}
\caption{Velocities of prominent emission lines in the 2019 November LBT spectrum. Most lines have been multiplied by a constant. The dashed gray line is the H$\alpha$ emission from the 2019 July Binospec spectrum, smoothed to match the resolution of MODS. }
\label{fig:component}
\end{figure}

All spectra exhibit prominent H$\alpha$ emission, but are otherwise almost featureless.  As Figure \ref{fig:halphacomp} shows, the H$\alpha$ emission line in all epochs appears to be multipeaked, with an absorption feature near $-20$\,km\,s$^{-1}$ as measured from our earliest spectrum on 2019 July 08 from the MMT. This spectrum can be reproduced by a combination of a broad Gaussian with full width at half-maximum intensity (FWHM) = 2000\,km\,s$^{-1}$ and a narrow Gaussian with FWHM = 350\,km\,s$^{-1}$, both in emission and centered at +135\,km\,s$^{-1}$, combined with a narrow Gaussian in absorption centered at $-20$\,km\,s$^{-1}$ and with FWHM = 115\,km\,s$^{-1}$.  The absorption feature is unresolved, so the model line-width is only an upper limit for the true FWHM of the absorption line. This P~Cygni absorption persists over the next few months, and as we show in Figure \ref{fig:component}, the absorption minimum and width are almost identical between the July MMT spectrum (dashed gray line) and the November LBT spectrum (red solid line). This indicates that the faster material from the eruption is still expanding into slower-moving CSM.

The strong \ion{Ca}{2} NIR triplet, which is not present in the progenitor, as well as \ion{Na}{1}~D $\lambda\lambda$5890, 5896 and very weak [\ion{Ca}{2}], are all seen in emission in the post-eruption spectra. A comparison of the prominent emission lines from our last epoch on 2019 November 01 is shown in Figure \ref{fig:component}, where some lines have been multiplied by a constant indicated in the legend for ease of viewing. While the red side of \ion{Ca}{2} NIR and H$\alpha$ are qualitatively similar, both lack an extended red shoulder that is seen in the other lines.  The absorption in the \ion{Ca}{2} lines at $-650$\,km\,s$^{-1}$, which is offset by $\sim 800$--900\,km\,s$^{-1}$ from the peak of the line, may indicate multiple locations for the various line emissions.  This could  be explained with an eruption in a dense, equatorial CSM, where the ejecta could expand much faster at the poles, yet slower in the plane of the disk where the [\ion{Ca}{2}] emission would arise.  A similar trend of faster H$\alpha$ and \ion{Ca}{2} and slower [\ion{Ca}{2}] was seen in the post-eruption spectra of UGC~2773-OT which may also have a bipolar nebula \citep{2016MNRAS.455.3546S}.

 \begin{table}
 \begin{center}
 \begin{minipage}{\linewidth}
\caption{Optical Spectroscopy of AT~2019krl}
\begin{tabular}{@{}lccccc}\hline\hline
Date & MJD & Telescope & R & Exp. \\
&&+Instrument&$\lambda$/$\Delta$$\lambda$&(s)\\
\hline
2018-11-13 &58435.41 & VLT+MUSE &2600 &3000\\
2019-07-08&58672.46&MMT+Binospec &3100&1800 \\
2019-07-09&58673.95&Lick Shane+Kast &770& 3600\\
2019-08-06&58701.35&SOAR+Goodman &1100&1800 \\
2019-08-28&58723.56&Keck+DEIMOS&1875&1200\\
2019-11-01&58788.19&LBT+MODS &2000&900 \\
\hline
\end{tabular}\label{tab:spectroscopy}
\end{minipage}
\end{center}
\end{table}

\section{Results} \label{Results}
\subsection{Constraints on the Progenitor}
The \textit{HST} and LBT data, along with the MUSE spectrum of the progenitor of AT~2019krl, allow us to thoroughly investigate the properties of the star that gave rise to this transient. Without a reliable value for the local extinction, our conservative choice of $E(B-V)=$ 0.12 mag will only provide lower limits to the mass and temperature of the progenitor, but will allow us to rule out certain classes of stars.

\begin{figure}
\includegraphics[width=\linewidth]{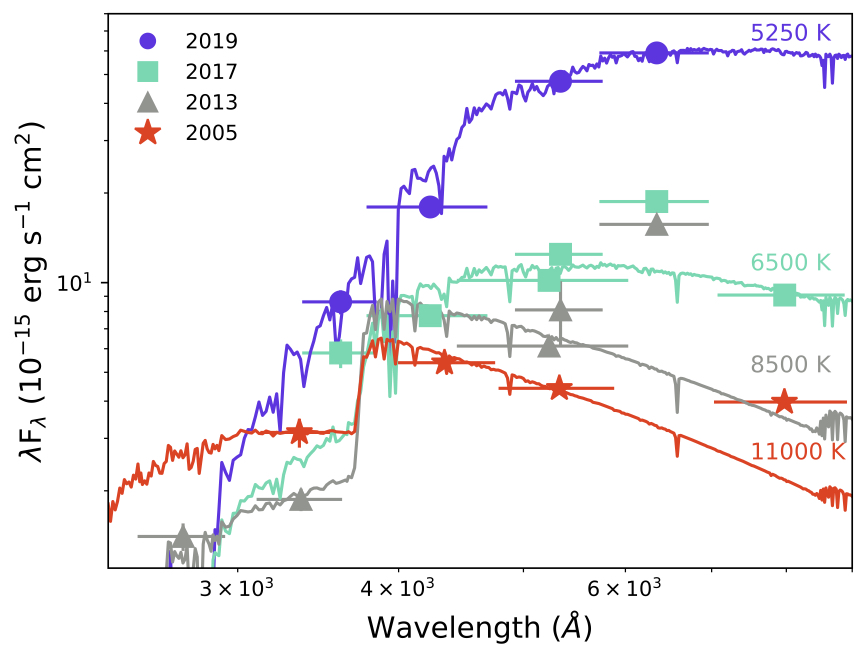}
\caption{Evolution of the SED of AT~2019krl from \textit{HST} (filled symbols) and LBT (open symbols) observations.  The data have only been corrected for $E(B-V) = 0.12$ mag. For comparison ATLAS synthetic spectra of stars with solar metallicity and log$(g)= 2.0$ are shown. }
\label{fig:SED}
\end{figure}

In Figure $\ref{fig:SED}$ we show the optical spectral energy distributions (SEDs) of the progenitor from photometry in 2005 (\textit{HST} only), 2013 (\textit{HST} and LBT), and 2017 (\textit{HST} and LBT). These epochs were chosen owing to the availability of the ultraviolet (UV) and $U$-band data, which provide the tightest constraints on the masses and temperatures of massive stars. We have attempted to fit the data with ATLAS synthetic spectra of stars of solar metallicity and log($g$) = 2.0 \citep{2003IAUS..210P.A20C}.  From all three epochs we can immediately rule out a cool progenitor, such as a red supergiant (RSG) or an AGB star, as even the minimum fit temperature of 6500 K is too high for those types of stars.  The P~Cygni absorption feature seen in the MMT spectrum is offset by $\sim 155$\,km\,s$^{-1}$ from the peak of H$\alpha$ traces the outflow wind velocity of the star, is also faster than typical RSG winds that have average wind velocities of 10--20\,km\,s$^{-1}$ \citep{MJ2011,Goldman17,Beasor18}.  Moreover, RSGs and AGB stars do not exhibit strong H$\alpha$ emission.

The 2005 epoch can be best fit by an 11,000\,K star with log($L$/L$_{\sun}$) = 4.4, although there is excess emission in the $F814W$ band that cannot be fit with just a single stellar model. In the subsequent two epochs the progenitor appears to cool and become more luminous with time, dropping to $T_{\rm eff}$ = 6500\,K with a higher luminosity of log($L$/L$_{\sun}$) = 4.6 by 2017. This is, of course, a lower limit, since any additional extinction (host or circumstellar) would raise both the temperature and luminosity. For instance, acceptable fits could be made to the 2005 data with a 17,000\,K model and $E(B-V) = 0.6$ mag. Note that we have not attempted to fit the $R$-band data in 2013 and 2017, as this filter contains the bright H$\alpha$ emission seen in the progenitor spectrum.

For comparison, yellow supergiants (YSGs) have  7500\,K $\gtrsim$ $T_{\rm eff}$ $\gtrsim$ 4800\,K and log($L$/L$_{\sun}$) $> 4.36$, with LBVs and blue supergiants (BSGs) exhibiting significantly warmer temperatures. LBVs in their cool outburst states typically have temperatures around 6000--10,000\,K.  The SED fits therefore indicate that the progenitor star was either a rather hot YSG, a quite cool BSG, or an LBV-like star in a cool phase. The $F555W-F814W$ color evolution (shown in Fig. \ref{fig:colorevol}) is too blue for an RSG, and is more consistent with a BSG or YSG. 
 
\begin{figure}
\begin{center}
\includegraphics[width=\linewidth]{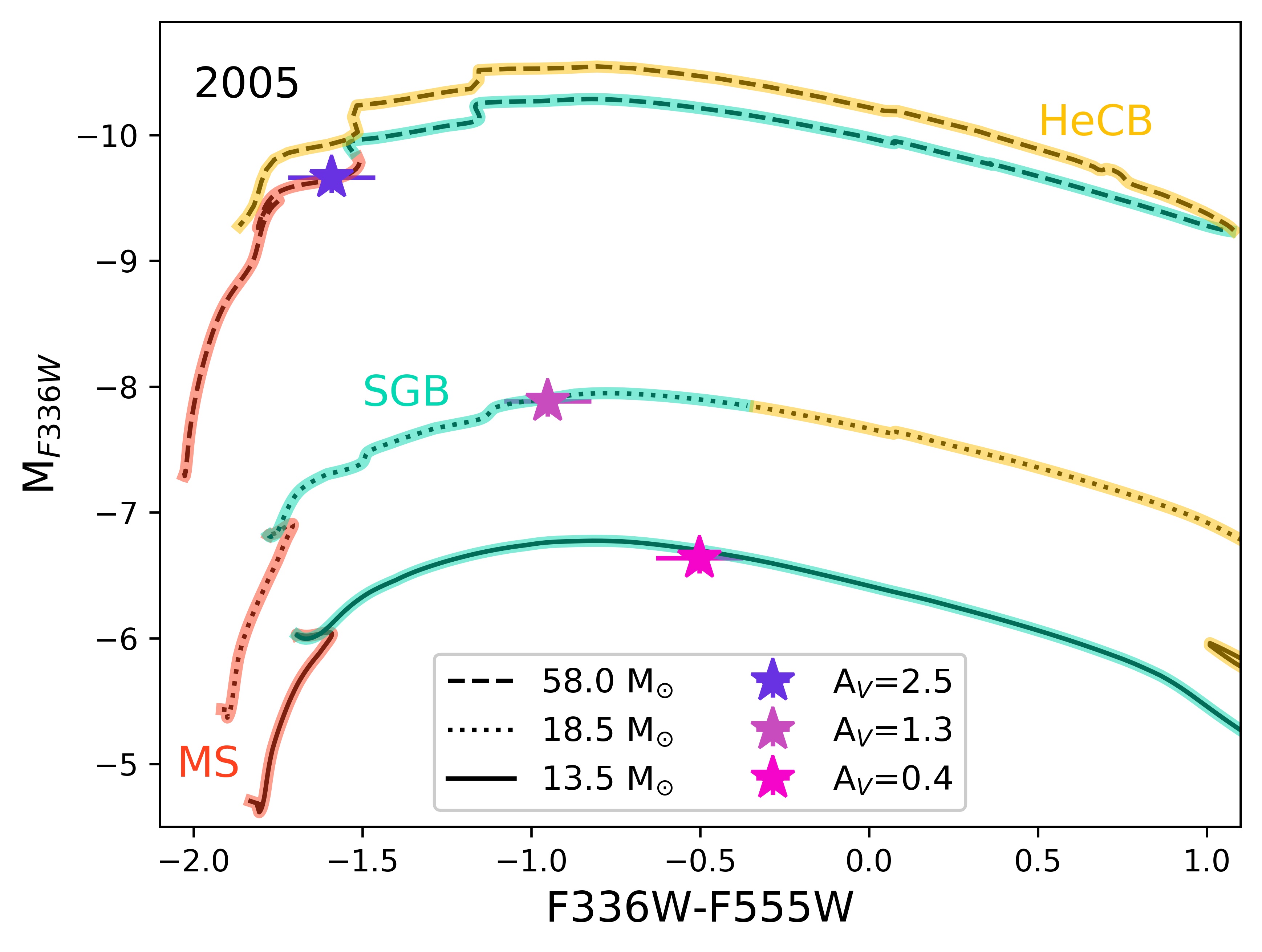}
\includegraphics[width=\linewidth]{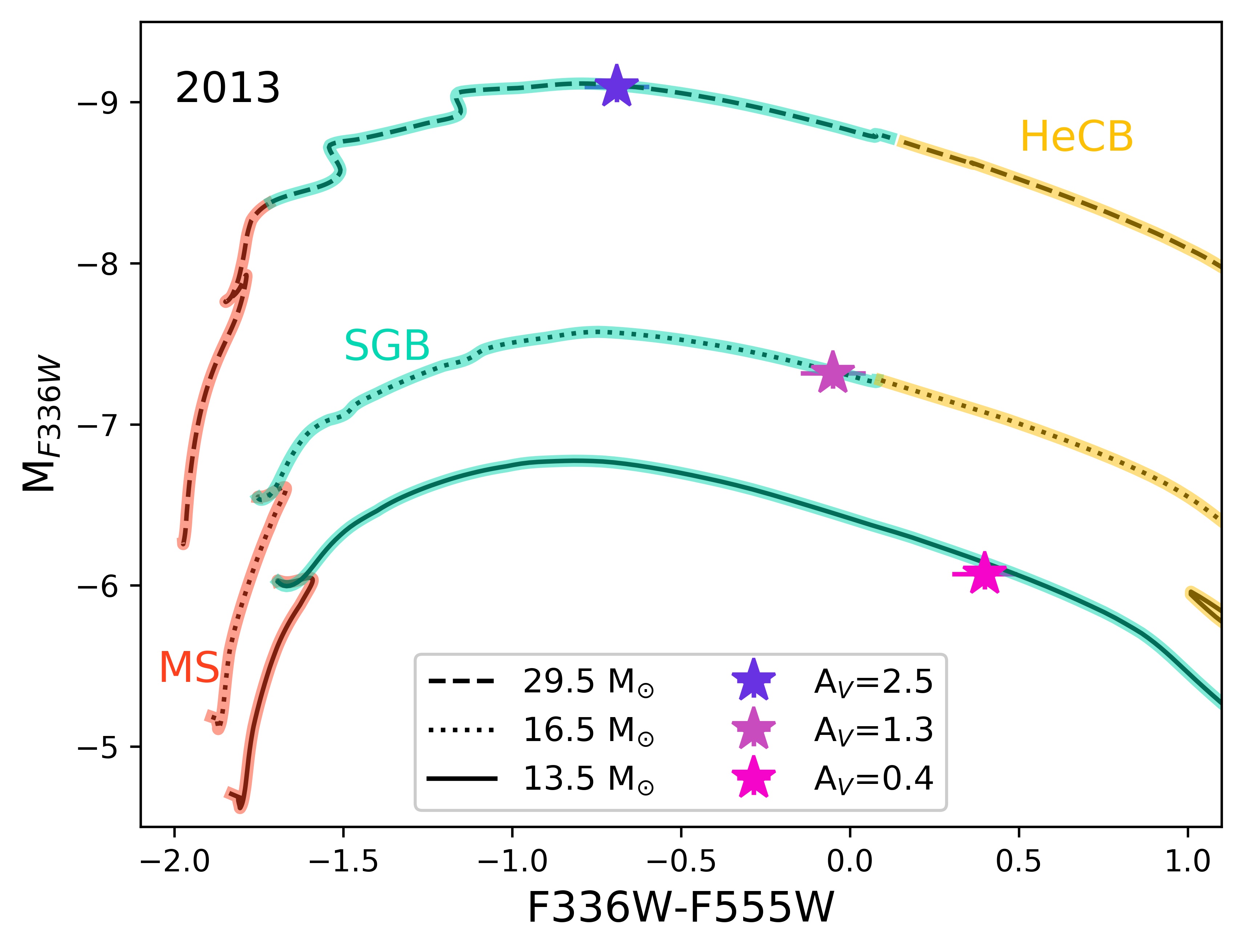}
\caption{Color-magnitude diagram of the progenitor of AT~2019krl (stars) in 2005 (top) and 2013 (bottom) with varying degrees of extinction applied. The main sequence (MS, red), supergiant branch (SGB, teal), and helium core burning (HeCB, yellow) are highlighted. The best-fit masses corresponding to each extinction value are shown as a solid, dotted, and dashed line, respectively. At no time was the SED as red as an AGB star or RSG.}
\label{fig:CMD2013}
\end{center}
\end{figure}

In Figure \ref{fig:CMD2013}  we compare the 2005 and 2013 \textit{HST} data to the MESA Isochrones $\&$ Stellar Tracks \citep[MIST\footnote{http://waps.cfa.harvard.edu/MIST/}]{2016ApJ...823..102C,2016ApJS..222....8D} to help constrain  the progenitor masses.  For each epoch we have determined the best-fit mass for three possible extinction values, with the lowest value of  $A_{V}$ = 0.4 mag corresponding to that used throughout this paper. The colors indicate the main sequence (MS, red), supergiant branch (SGB, teal), and helium core burning (HeCB, yellow)  phases, and the solid, dotted, and dashed lines show the various mass tracks. In both 2005 and 2013 we find a lower limit to the progenitor mass of 13.5 M$_{\sun}$; however, the data with the largest amount of extinction applied yield a progenitor mass of 58 M$_{\sun}$ in 2005 and only 29.5 M$_{\sun}$ in 2013.

\begin{figure}
\begin{center}
\includegraphics[width=\linewidth]{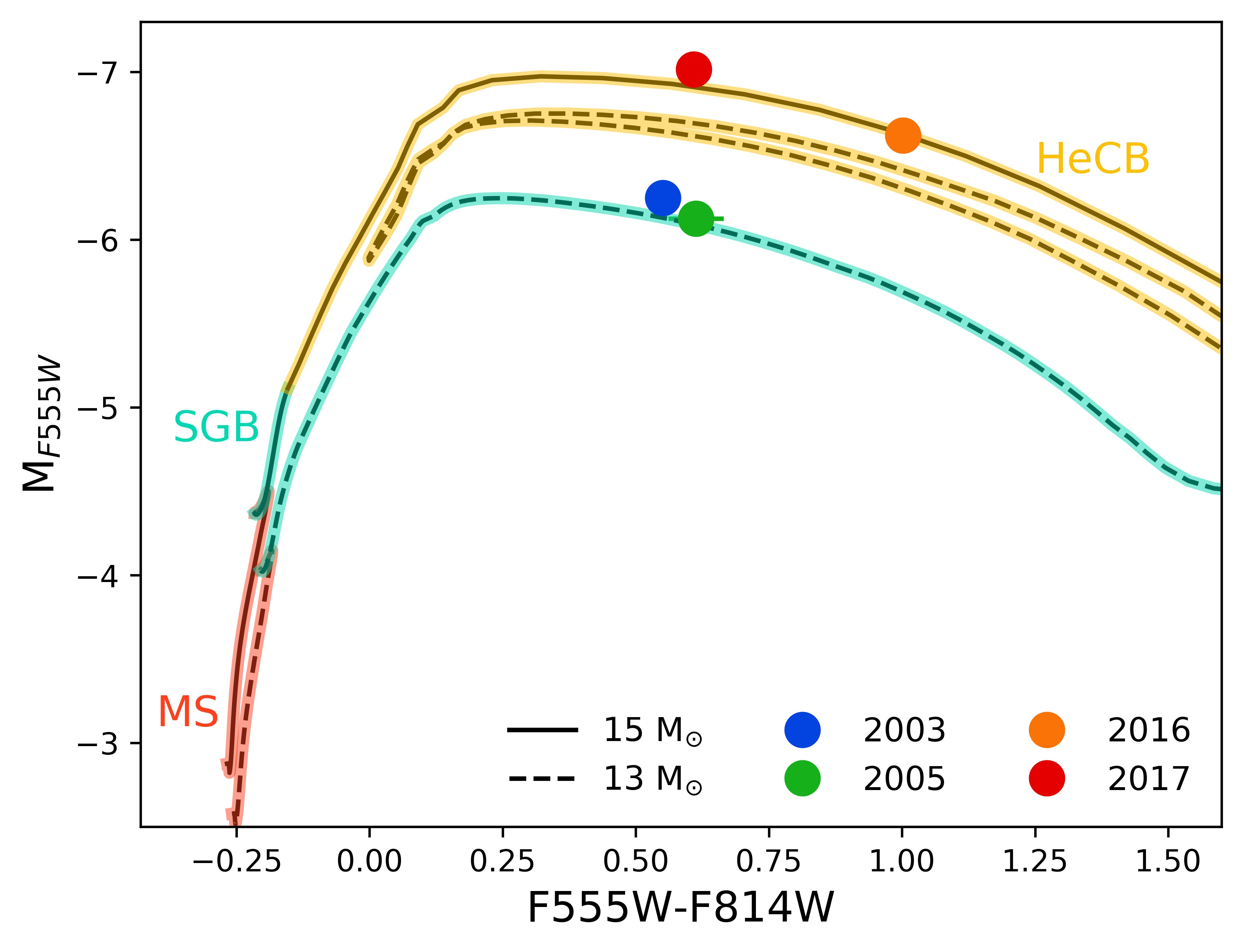}
\caption{Same as Figure \ref{fig:CMD2013} but illustrating the evolution of the progenitor of AT~2019krl within the $F555W-F814W$ color-magnitude diagram with time, corrected for $E(B-V)$=0.12 mag.}
\label{fig:CMD555814}
\end{center}
\end{figure} 

To illustrate how the progenitor mass estimate changes depending on the epoch, we show the evolution of the source in $M_{F555W}$ and  $F555W-F814W$ in Figure \ref{fig:CMD555814}.  Similar to the SED fits, there is a trend to redder colors with time.  This translates to shifts in progenitor mass estimates from around 13 M$_{\sun}$ to 15 M$_{\sun}$, but also a shift in the inferred evolutionary stage from SGB to HeCB. This, of course, is not real evolution, as the change to helium core burning takes significantly more time than a mere 15\,yr.  Instead, it illustrates how changes in the stellar structure due to instability before an eruption can mimic observed evolutionary changes; values for the inferred mass or luminosity from any single epoch of such a transient should therefore be regarded with caution. 

\begin{figure}
\includegraphics[width=\linewidth]{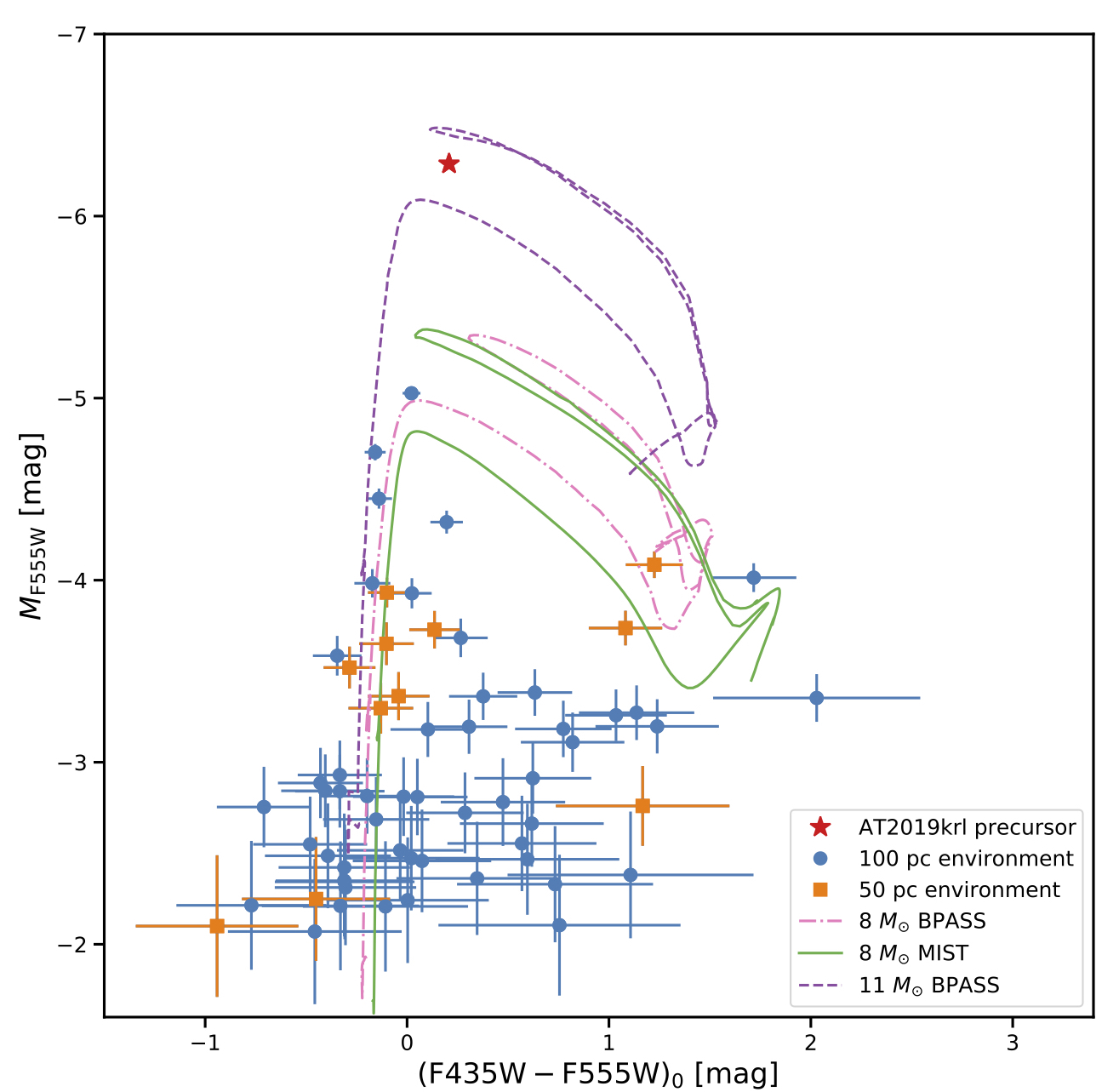}
\includegraphics[width=\linewidth]{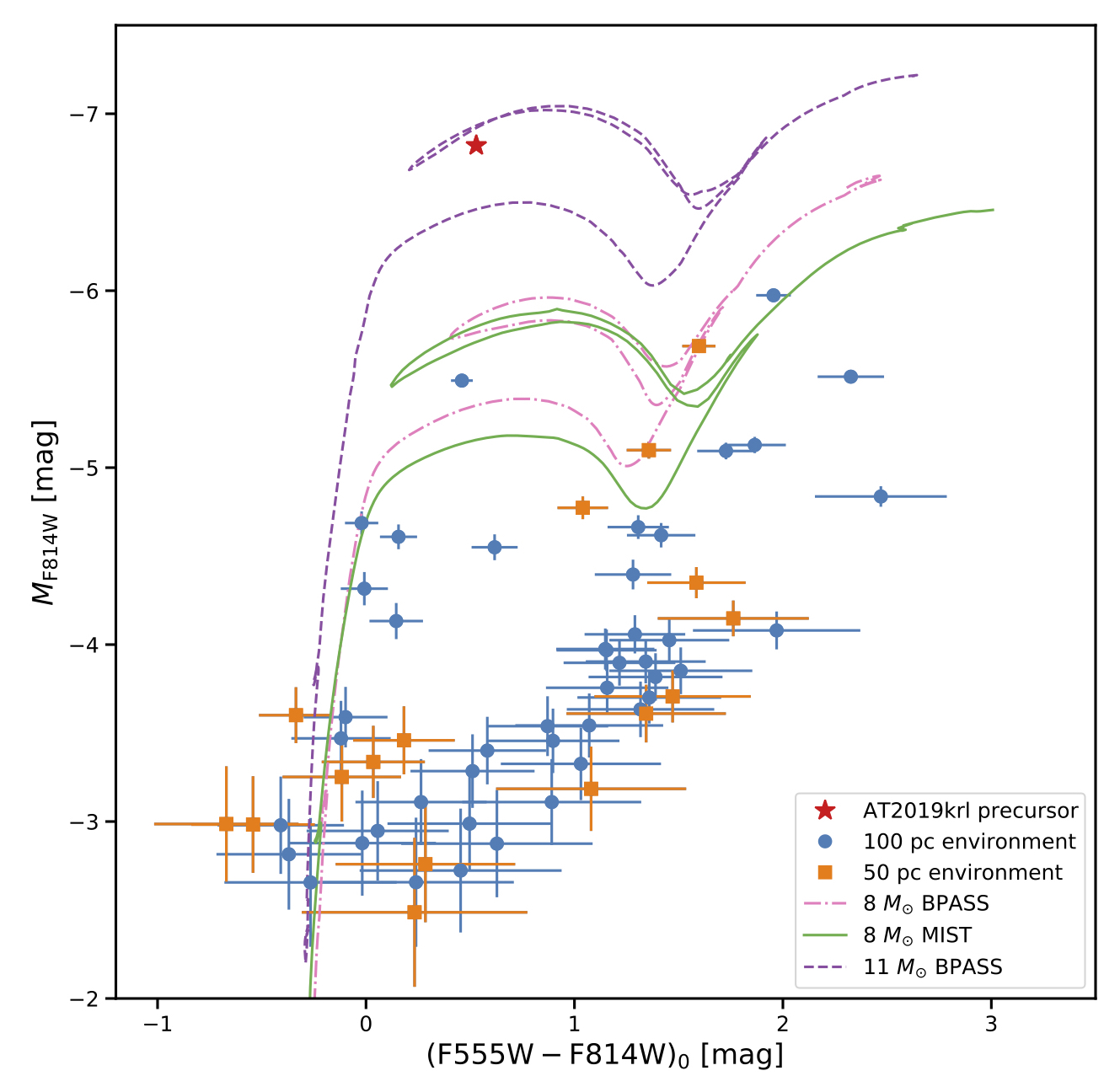}
\caption{Color-magnitude diagrams of the immediate environment around AT~2019krl from the 2003 \textit{HST} data, adjusted to the assumed distance and reddening. We have isolated the stars within 50 pc and 100 pc of the transient \citep[see, e.g., ][]{Williams2018,Schady2019}, which is shown with the star symbol (the photometric uncertainties for AT~2019krl are smaller than the symbol). Also shown for comparison are theoretical evolutionary tracks at solar metallicity for single stars at 11 and 8 M$_\sun$ from BPASS \citep{Stanway2018} and 8 M$_\sun$ from MIST \citep{2016ApJ...823..102C,2016ApJS..222....8D}.}
\label{fig:CMDprog}
\end{figure}

We can also use the stellar population of the local environment surrounding AT~2019krl to put some constraints on the progenitor. Color-magnitude diagrams assembled from the 2003 \textit{HST} data, assuming $E(B-V) = 0.12$ mag and shown in Figure \ref{fig:CMDprog}, reveal no stars brighter than $M_{F555W}$ = --4 mag and $M_{F814W}$ = --6 mag within 50\,pc of the progenitor star, and none brighter than $M_{F555W}$ = --5 mag within 100\,pc. This nominally suggests a lack of stars more massive than 8\,M$_{\sun}$ surrounding AT~2019krl, and that the progenitor of AT~2019krl is overluminous for the age that would be inferred from the surrounding  stars.  One way to create this scenario is through a binary rejuvenation in a blue-straggler star \citep{2015MNRAS.447..598S}. 

\begin{figure}
\includegraphics[width=\linewidth]{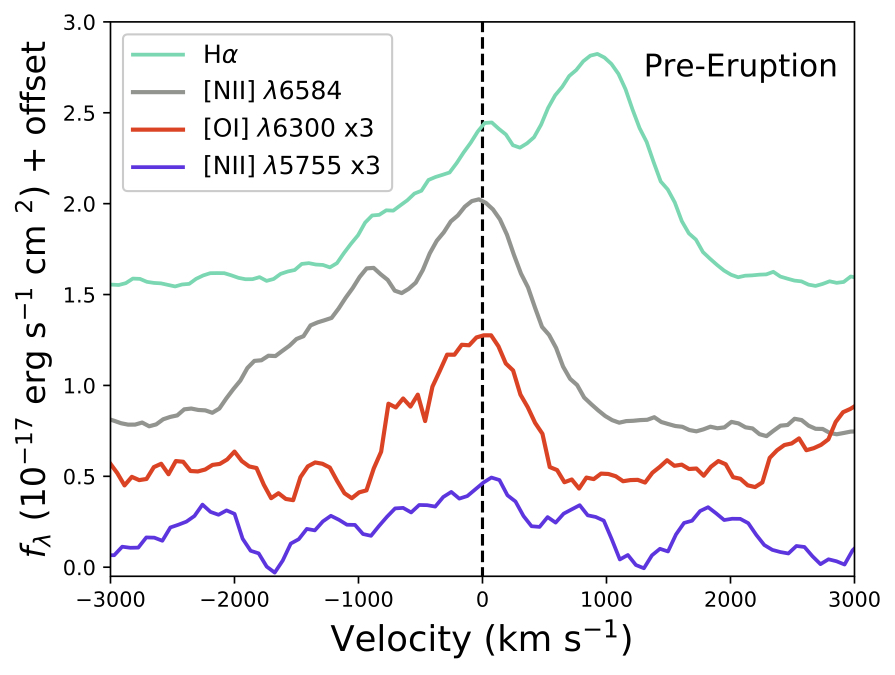}
\includegraphics[width=\linewidth]{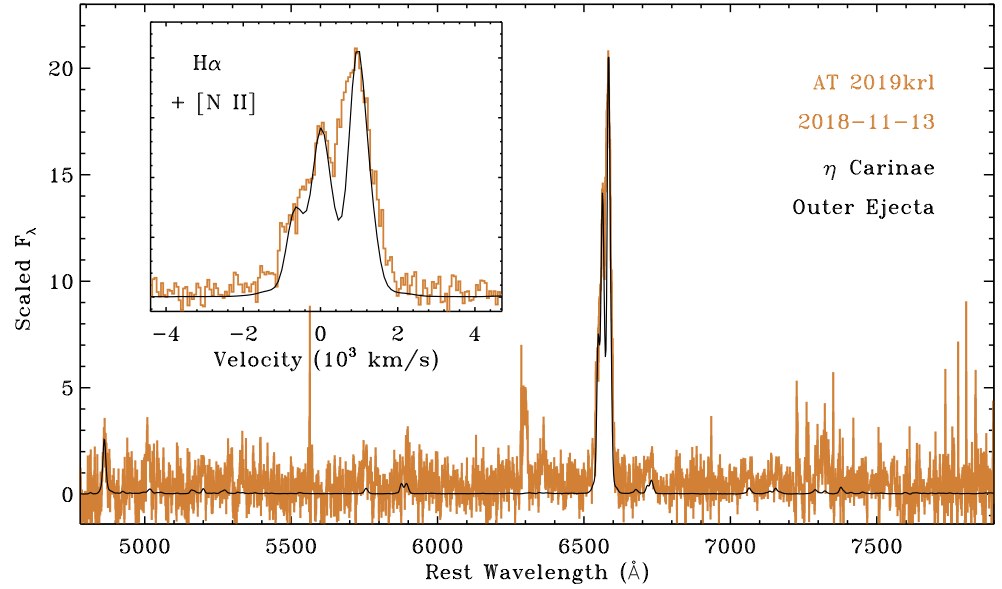}
\caption{(Top:) Velocities of prominent emission lines in the 2018 pre-eruption spectrum. Most lines have been multiplied by a constant indicated in the legend and have been smoothed for ease of viewing. (Bottom:) Pre-eruption 2018 MUSE spectrum (orange) compared with the S~Ridge in the outer ejecta of $\eta$~Car. The $\eta$~Car spectrum has been scaled to match the [\ion{N}{2}] $\lambda$6583 emission.} 
\label{fig:preerupspec}
\end{figure}

Finally, the progenitor spectrum provides clues about the physical state of the star prior to eruption. The spectrum is dominated by broad H$\alpha$ with wings extending to roughly $\pm$ 2000\,km\,s$^{-1}$ (Fig. \ref{fig:spectra}).  Strong [\ion{N}{2}] $\lambda$6584 emission is present as well,  much stronger than the H$\alpha$ emission.  Unlike the narrow emission lines of nearby H~II regions that have FWHM at the resolution limit of the spectrum of $\sim 115$\,km\,s$^{-1}$, the [\ion{N}{2}] $\lambda$5755, $\lambda$6584 and [\ion{O}{1}] $\lambda\lambda$6300, 6363 lines have much broader FWHM as shown in Figure \ref{fig:preerupspec} and listed in Table \ref{tab:progspec}.  This may point to emission of N-rich CSM, commonly seen around massive stars, although the excitation of the [\ion{N}{2}] emission is uncertain and may be complicated by a mix of shock excitation and photoionization. In particular, shock excitation cannot be ignored as the broad line width is much faster than typical BSG or cool supergiant wind speeds. 

The high [\ion{N}{2}]/H$\alpha$ intensity ratio and the width of almost $10^{3}$\,km\,s$^{-1}$ are reminiscent of the Outer Ejecta of $\eta$ Car \citep{2004ApJ...605..854S}.  In Figure \ref{fig:preerupspec} we show the pre-explosion progenitor spectrum of AT~2019krl compared to that of the S~Ridge in the outer ejecta of $\eta$~Car from \citet{2004ApJ...605..854S}, which has been scaled to the [\ion{N}{2}] $\lambda$6584 line strength of AT~2019krl. There are striking similarities between the two spectra. In the case of $\eta$ Car, the high [\ion{N}{2}]/H$\alpha$ ratio arises in very N-rich CSM ejected several hundred years prior to the main eruption, with expansion speeds faster than the bulk outflow in the main eruption or the present-day wind \citep{kiminki16}.   In these ejecta around $\eta$ Car, the emission is powered by shock excitation as very fast ejecta overtake the CSM \citep{smith08,2004ApJ...605..854S} and these N-rich ejecta are seen alongside a bright soft X-ray shell \citep{2001ApJ...553..832S}. There may also be photoionization from O-type stars in its surroundings, but the central star does not photoionize these ejecta, because they reside outside thick layers of CSM with neutral atomic gas, molecular gas, and dust \citep{sm19}.

It is plausible that the same mechanisms responsible for the N-rich emission seen in $\eta$ Car are at play in AT~2019krl, although the data are far less constraining for AT~2019krl. Regardless of  excitation and chemical abundance, the strong, broad [\ion{N}{2}] emission does point to prior episodes of mass loss with speeds faster than the progenitor's wind, and that the current eruption is plowing into material lost in a previous eruption. Any previous eruption would have had to happen prior to 2003 or it would have been detected in our light-curve data.  Therefore, we can estimate a lower limit of a radius of the nebula surrounding AT~2019krl to be 1100 km s$^{-1}$ $\times$ 16\,yr, or $5.5 \times 10^{16}$\,cm. 

Photoionization may still contribute to the [\ion{N}{2}] emission in AT~2019krl, although as we discuss below, the lack of nearby O-type stars combined with the fast [\ion{N}{2}] may make this scenario less likely.
It is also possible that other lines of ionized N may be present, but are lost in the noise of the spectrum, and the [\ion{N}{2}] emission strength may be unrelated to the progenitor star's temperature. The higher-excitation lines seen in the spectrum of $\eta$~Car's ejecta that clearly require shock excitation are below the noise level in our progenitor spectrum of AT~2019krl.

Another class of stars that could possibly show this level of H$\alpha$ flux are sgB[e] stars, which are easily confused with LBVs in quiescence, since they can appear spectroscopically similar, and have similar temperatures and luminosities. The sgB[e] stars generally show [\ion{O}{1}] emission \citep{2016MNRAS.456.1424A} which is not seen in the post-eruption spectra of AT 2019krl, although it is present in the pre-eruption spectrum with a simlar width to the [\ion{N}{2}] lines (Fig. \ref{fig:preerupspec}). If the [\ion{O}{1}] and [\ion{N}{2}] emission are due to the B[e] phenomenon, their high velocities ($\sim 1000$\,km\,s$^{-1}$) are puzzling, since velocities in sgB[e] stars are generally on the order of 200--300\,km\,s$^{-1}$, with higher velocities confined to the electron-scattering wings in H$\alpha$, if present \citep[for example]{2013A&A...560A..11C,2016MNRAS.456.1424A,2018A&A...612A.113T}. 

\subsection{Post-Eruption Analysis}
The light-curve peak was unfortunately missed in the optical. However, we can use the change in magnitude in the \textit{Spitzer} MIR fluxes to estimate an approximate peak magnitude in the optical light curve bands.  The 4.5\,$\mu$m data increased by 6.75 mag between 2018 November and the peak in 2019 May.  If we assume a similar change in the $R$-band luminosity from 2018 November, then the peak would be $M_{R} = -$14.6 mag. If instead we assume that the color difference between $R$ and other bands remains the same at peak outburst as in late 2017, we can estimate the maximum absolute brightness of $M_{F814W} = -$14.3 mag, $M_{V} = -$13.8 mag, and $M_{F555W} = -$13.5 mag on 2019 May 17. This is well within the distribution of peak visual-wavelength absolute magnitudes for other SN impostors and/or giant LBV eruptions \citep{2011MNRAS.415..773S}.

Comparison of the H$\alpha$ evolution (Figure \ref{fig:spectra}) also shows very little change in the H$\alpha$ line profile from our first spectrum, $\sim 60$ days after eruption, to the last epoch on $\sim$ 180 days.  Additionally,  the presence of weak [\ion{Ca}{2}] yet relatively strong \ion{Ca}{2} emission may provide some insight into the circumstellar environment of AT~2019krl. We can use the ratio of these lines to estimate the electron densities using the prescription of \citet{2013ApJ...773...46H}. From the 2019 November spectrum we obtain an estimate of $n_e = 1.1 \times 10^{7}$\,cm$^{-3}$.  This of course assumes that the emission lines are coming from the same region, which may not be accurate,  since the two sets of lines exhibit different temporal evolution and different line profiles.  In SN~2008S \citep{Prieto2008S,2009ApJ...697L..49S} and  UGC~2773-OT \citep{2010AJ....139.1451S}, these forbidden emission lines were linked to vaporizing dust in the CSM during the outburst.  The same may have occurred in AT~2019krl, as dust grains that formed around the progenitor may have evaporated during the sudden luminosity increase of the eruption \citep{2011ApJ...743...73K}.

\section{Discussion} \label{discussion}

\begin{figure*}
\begin{center}
\includegraphics[width=3.5in]{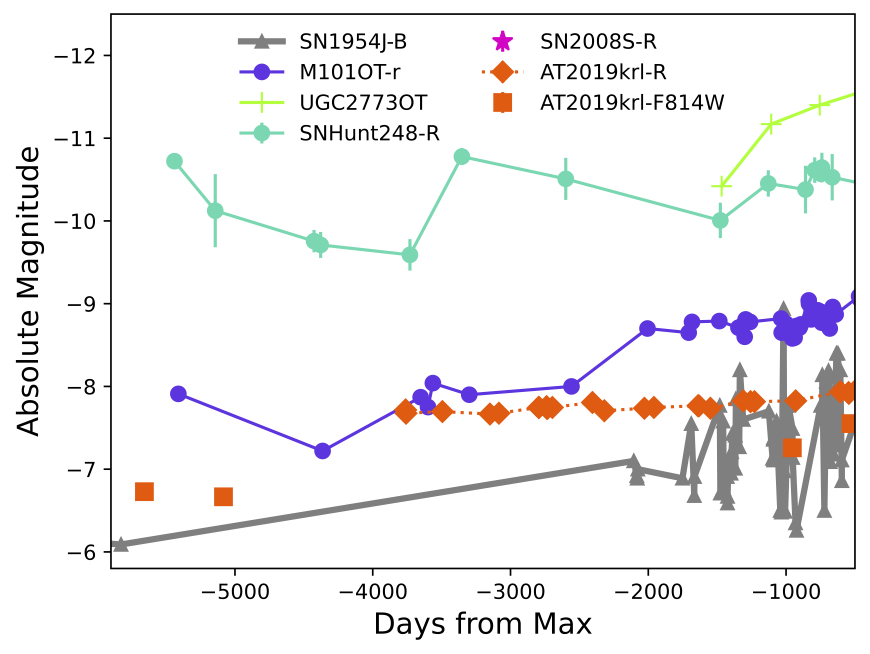}
\includegraphics[width=3.55in]{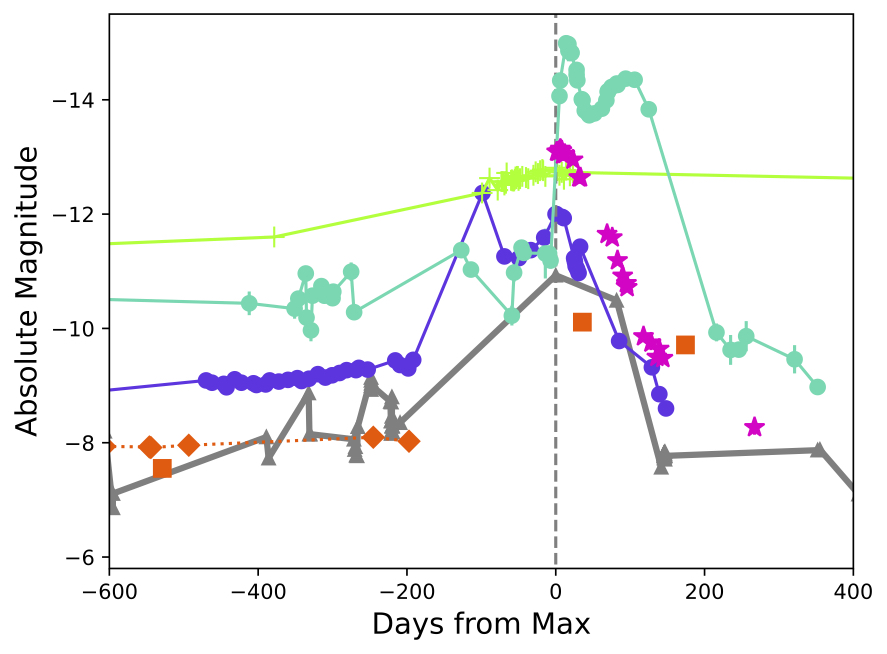}
\caption{Comparison of the $R$ and $F814W$ light curves of AT~2019krl to other transients. The left panel shows the historical light curves of the progenitors, while the right panel focuses on the eruption.  All data have been corrected for their respective $E(B-V)$.  Data are from \citet[M101 OT2015-1]{2017ApJ...834..107B}, \citet[SNhunt248]{2015A&A...581L...4K}, \citet[SN1954J]{1968ApJ...151..825T}, \citet[SN~2008S]{Botticella2008S}, and \citet[UGC2773-OT]{2010AJ....139.1451S,2016MNRAS.455.3546S}. Note that the UGC2773-OT data are unfiltered.}
\label{fig:LCcomp}
\end{center}
\end{figure*}

Even though the optical peak of the outburst was missed because AT~2019krl was behind the Sun, the extensive data on the pre-eruption star provides new and important clues into the progenitor.
The observational properties overlap significantly with other transients, and below we  discuss how AT~2019krl is like and unlike various transient event classes.

\subsection{Comparison to LBV eruptions}
While LBVs can experience low-amplitude, irregular, S-Doradus variations, where the peak of the luminosity is thought to shift from the UV to the optical and the star brightens $\sim$ 1--2 mag, they can also go through a rare form of eruptive mass loss referred to as giant eruptions.  Many of the so-called ``SN imposters" have been interpreted as these giant eruptions of LBVs, similar to the historical eruption of P~Cygni or the Great Eruption of $\eta$~Car \citep{2000PASP..112.1532V,2011MNRAS.415..773S}. During these eruptions the luminosity of the star increases while the temperature usually drops. The eruptive phase of an LBV can last for years, as in the cases of $\eta$ Car \citep{2011MNRAS.415.2009S} and UGC~2773-OT \citep[Figure \ref{fig:LCcomp}]{2016MNRAS.455.3546S}. Additionally, quiescent or eruptive LBV winds can lie in the 100--600\,km\,s$^{-1}$ range, similar to the resolved narrow H$\alpha$ component seen in AT~2019krl. The overall appearance of the spectrum in AT~2019krl --- including the [\ion{Ca}{2}] and \ion{Ca}{2} lines, along with the comparable H$\alpha$ profiles and line strengths, the inferred temperature, the dusty CSM, and the IR excess --- are all consistent with known LBV giant eruptions. 

In Figure \ref{fig:LCcomp} we show the $B$-band light curve of SN~1954J \citep{1968ApJ...151..825T}, thought to be the eruption of a luminous ($M_{V}\approx -$8.0 mag) and massive ($>$ 25 M$_{\sun}$)  LBV  \citep{2005PASP..117..553V}. The light curves appear similar, except for the small amplitude variability seen in SN~1954J, which is on a fast enough timescale to have been missed by the cadence of the progenitor data for AT~2019krl. We also show  the unfiltered light curve of UGC~2773-OT in Figure \ref{fig:LCcomp}. Unlike AT~2019krl, UGC~2773-OT had a much more gradual rise to brightness, but both events exhibit a slow decline in luminosity post-peak. Estimates for the mass of UGC~2773-OT are $\sim$ 20\,M$_{\sun}$ (or greater if larger extinction is adopted), which is similar to that of AT~2019krl \citep{2010AJ....139.1451S,2011ApJ...732...32F}, and both have inferred dusty, asymmetric CSM.   

Although the masses and luminosities estimated above for AT~2019krl, using the modest extinction of $A_{V} = 0.4$ mag, are significantly lower than those traditionally associated with LBVs \citep{2011MNRAS.415..773S}, only an additional 1--1.5 mag of extinction could easily push AT~2019krl's progenitor to higher masses, as shown in Figure \ref{fig:CMD2013}. Moreover, recent studies with revised distances have shown that Milky Way LBVs extend to lower initial masses and luminosities than previously thought \citep{2019MNRAS.488.1760S}. The brightness of the eruption and the slow evolution afterward, combined with the color and mass of the progenitor, provide strong evidence for the possible LBV-like nature of AT~2019krl.

\subsection{Comparison to SN~2008S-Type Events}
One class of transients with progenitors that are very bright in the IR are the SN~2008S-like events. Well-studied members of this class include the namesake SN~2008S \citep{Prieto2008S,Botticella2008S,2009ApJ...697L..49S}, NGC~300 OT2008-1 \citep{2009ApJ...699.1850B,2009ApJ...695L.154B,2011ApJ...743..118H,2009ApJ...705.1364T,2012ApJ...758..142K}, SN~2002bu \citep{2011MNRAS.415..773S,2012ApJ...760...20S}, PTF10fqs \citep{2011ApJ...730..134K},  AT2017be \citep{2018MNRAS.480.3424C}, and M51 OT2019-1 \citep{Jencson2019,2020A&A...637A..20W}. 

\begin{figure*}
\includegraphics[width=3.5in, height=2.6in]{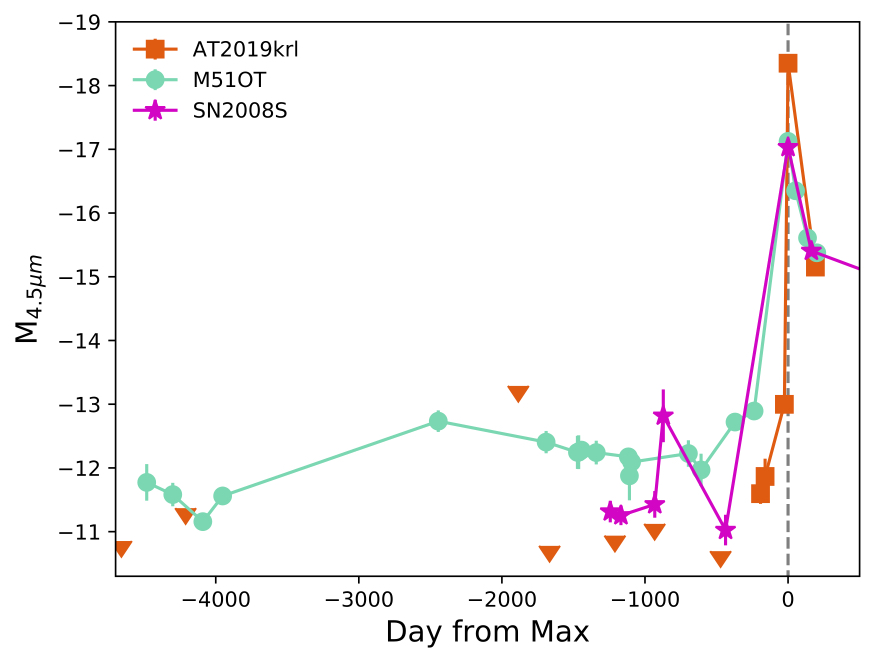}
\includegraphics[width=3.3in]{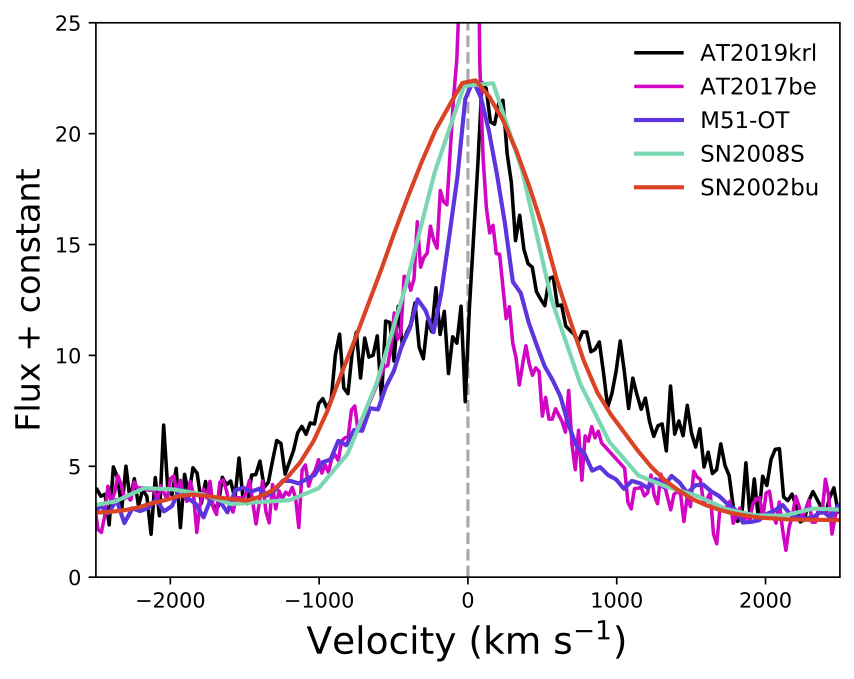}
\caption{Left: $Spitzer$ 4.5\,$\mu$m light curves of the progenitors and eruptions of AT~2019krl, M51OT \citep{Jencson2019}, and SN~2008S \citep{Prieto2008S}. For SN~2008S we are using the distance derived to NGC 6946 in \citet{NGC6946TGRB} of 7.72 Mpc. Right: H$\alpha$ emission line profiles of SN~2008S type events at around 60--70 days.  The AT~2019krl spectrum is from 2019 July 08, the AT~2017be spectrum is unpublished from MMT/Bluechannel taken on 2017 March 08, M51OT is from 2019 March 07 \citep{Jencson2019}, SN~2008S is from 2008 April 25 \citep{Botticella2008S}, and SN~2002bu from 2002 June 08 \citep{2011MNRAS.415..773S}. }
\label{fig:IRLT}
\end{figure*}

The SN~2008S-type transients show strong Balmer, \ion{Ca}{2} NIR triplet, and [\ion{Ca}{2}] emission in their spectra, with outflow velocities on the order of 500--1000\,km\,s$^{-1}$, similar to many LBV great eruptions. In Figure \ref{fig:IRLT} we show the H$\alpha$ emission at $\sim$70 days post-peak for a sample of SN~2008S-like events.  All exhibit fairly smooth profiles with wings extending to $\sim 1000$\,km\,s$^{-1}$; however, for AT~2019krl the emission line is broader and multipeaked. While AT~2019krl shows  strong H$\alpha$ and the \ion{Ca}{2} NIR triplet, the [\ion{Ca}{2}] emission is quite weak and appears months after peak. This is unlike SN 2008S-like events which have strong, prompt [\ion{Ca}{2}] emission. 

The photometric evolution is also dissimilar between the SN~2008S-type events and AT~2019krl, as shown in Figures \ref{fig:LCcomp} and \ref{fig:IRLT}.  In particular, AT~2019krl is at least 1.5\,mag brighter in the  4.5\,$\mu$m band at peak, and while we are unsure of its brightness in the $R$ band at peak, we do know that it fades quite slowly and is brighter than SN~2008S by day 200. Furthermore, the late-time evolution of SN~2008S-type events seems to fade well below the luminosity of the progenitor, particularly in the IR \citep{2016MNRAS.460.1645A}. This of course is expected for terminal events. We do not currently have the post-eruption observation to tell if this is the case for AT~2019krl, so continued observations are needed. 

This class of transients has been associated with highly dust-obscured progenitors that only appear in the MIR \citep{Prieto2008S,2009ApJ...705.1364T}, and often show signatures of dust in the months following eruption, either surviving or newly formed \citep{2009ApJ...705.1425P}. Of course, if the dusty CSM has a nonspherical geometry, then the amount of dust obscuration for the progenitor may vary widely depending on viewing angle \citep{2009ApJ...697L..49S,2011MNRAS.415..773S, 2020arXiv201014490S}. An asymmetric CSM around NGC~300-OT was proposed based on optical spectropolarimetry \citep{2010A&A...510A.108P} and NIR spectroscopy \citep{2010ApJ...718.1456O}. If the same is true for AT~2019krl, and the dust is confined to a torus around the progenitor star which we happen to view pole-on, it could appear bright in both the optical and IR.

\begin{figure}
\includegraphics[width=\linewidth]{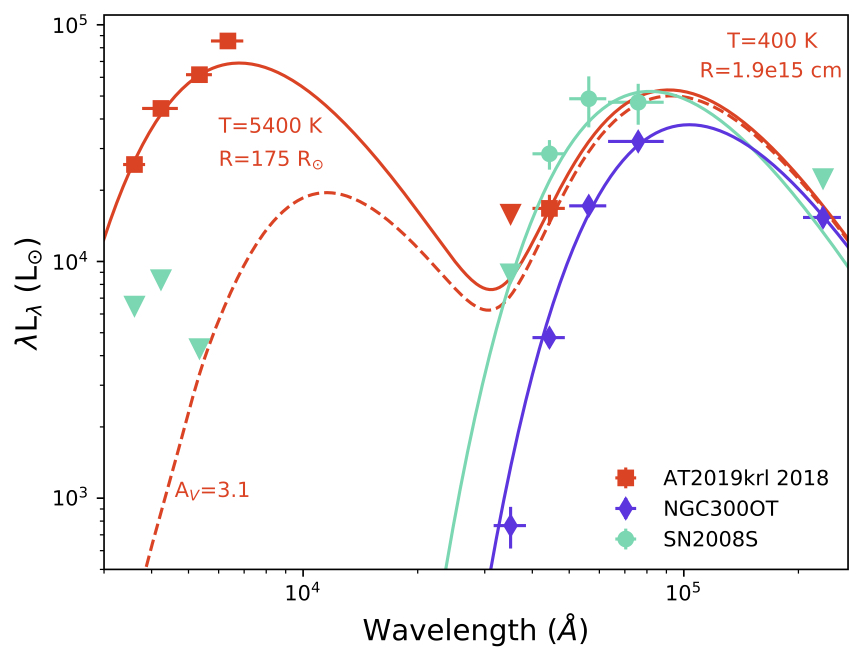}
\caption{SED of the progenitor of AT~2019krl compared to the progenitors of SN~2008S \citep{Prieto2008S} and NGC~300-OT \citep{2009ApJ...705.1425P}. The MIR of AT~2019krl has been fit assuming a 400\,K blackbody as described in the text. The dashed red line shows the combined blackbody fit to the AT~2019krl data (solid line) that has been reddened with $E(B-V) = 1.0$ mag. }
\label{fig:SEDcomp}
\end{figure}

In Figure \ref{fig:SEDcomp} we compare the optical and MIR SEDs for the progenitor of AT~2019krl to the progenitor of SN~20008S and NGC~300-OT.  The detection of the optical component only yielded upper limits for SN~2008S and NGC~300-OT, while in every epoch of the AT~2019krl progenitor, we have significant detections. The much closer NGC~300-OT had clear progenitor detections in all of the IRAC bands, while SN~2008S and AT~2019krl had only an upper limit estimated from the 3.6\,$\mu$m images.  The 2018 \textit{Spitzer} data for AT~2019krl are from the post-cryogenic mission only, so there is no way to determine the brightness in the 5.8\,$\mu$m and 8.0\,$\mu$m bands, but we have attempted to fit an MIR component with a blackbody temperature of 400\,K, between the two temperature values of SN~2008S \citep[440K]{Prieto2008S,2009ApJ...705.1364T} and NGC~300-OT \citep[338K]{2009ApJ...699.1850B}. As we mention above, upper limits were measured for 5.8\,$\mu$m and 8.0\,$\mu$m during the cryogenic mission, but the increase in the MIR luminosity of AT~2019krl by 2018 suggests that it would have been detected in these longer wavelength bands.

\begin{table}
 \begin{center}
 \begin{minipage}{\linewidth}
\caption{Progenitor spectrum emission line properties}
\begin{tabular}{@{}ccccc}\hline\hline
Line & EW & Flux & FWHM & FWHM \\
\AA&\AA&10$^{-17}$ erg s$^{-1}$ cm$^{2}$&\AA& km s$^{-1}$\\
\hline
5755&28 $\pm$ 9 &2.0 $\pm$ 0.3 &19 $\pm$ 3&970 $\pm$ 50\\
6300&72 $\pm$ 10&4.6 $\pm$ 0.3&15 $\pm$ 1&770 $\pm$ 50\\
6548&194 $\pm$ 20&14.2 $\pm$ 0.8&25 $\pm$ 1&1180 $\pm$ 100\\
6563&46 $\pm$ 13&3.4 $\pm$ 0.7&8 $\pm$ 1&350 $\pm$ 30\\
6584&397 $\pm$ 35&29.0 $\pm$ 1.0&22 $\pm$ 1&1000 $\pm$ 100\\
\hline
\end{tabular}\label{tab:progspec}
\end{minipage}
\end{center}
\end{table}

NGC~300-OT was also detected in the 24\,$\mu$m MIPS band, while only an upper limit could be derived for SN~2008S. It is more ambiguous in the case of AT~2019krl, since a detection was made at the location of the progenitor in the MIPS 24\,$\mu$m data in 2005, but the mitigating factors of pixel size and the distance of M74 make it difficult to determine if the flux originates from the transient, as opposed to distant surrounding material associated with star formation.  If the 24\,$\mu$m flux comes from AT~2019krl, then it suggests a second region of cooler dust much further out than the warmer dust mapped by the 3.6\,$\mu$m and 4.5\,$\mu$m fluxes. 
Given the uncertain origin of the 24\,$\mu$m flux and the fact that IRAC data at other MIR wavelengths gave only upper limits, we cannot provide good constraints on a unique fit for this cool component. 
 
After the discovery of SN~2008S and NGC~300-OT, \citet{2009ApJ...705.1364T} suggested that they constitute a new class of transients that may be caused by ecSNe, an explanation also suggested by \cite{Botticella2008S}. The explosion of a super-AGB (sAGB) star as an ecSN has an expected kinetic energy of $\sim 10^{50}$ erg, and progenitors are thought to be in the initial mass range 8--10 M$_{\sun}$.  The exact mass range is still debated, and may be very narrow \citep{2015MNRAS.446.2599D}. The progenitor photometry for AT~2019krl points to the equivalent of a single star initially more massive than 13 M$_{\sun}$. Even with no extinction correction, it was much hotter and likely less dust-enshrouded than an sAGB star. This clearly rules out an ecSN from a sAGB star for the case of AT~2019krl. Additionally, sAGB stars pulsate with large variability ($> 1$ mag) in their light curves, particularly in the IR \citep{2009ApJ...705.1364T}. These variations are not seen in the progenitor of AT~2019krl (Figure \ref{fig:IRLT}), at least to brightness levels that would be above the detection limit of the available \textit{Spitzer} observations.

Plausible alternative progenitor scenarios to this class of events are the outburst of a heavily obscured LBV \citep{2009ApJ...697L..49S,2011MNRAS.415..773S}, or other dust enshrouded massive star in a binary system \citep{2009ApJ...699.1850B,2009ApJ...695L.154B,2011MNRAS.415..773S}. SN~2008S had an estimated total extinction of $A_{V} = 2.5$ mag at peak \citep{Prieto2008S} and M51-OT a total reddening $0.7 < E(B-V) {\rm [mag]} <$ 0.9 \citep{Jencson2019}, which for $R_{V} =3.1$ translates to  $2.2 < A_{V} < 2.8$ mag.  If we assume a total $A_{V} = 2.5$ mag for AT~2019krl, then the best fit implies a stellar mass as high as 58 M$_{\sun}$ from the 2005 \textit{HST} data, or 29.5 M$_{\sun}$ from the 2013 \textit{HST} data, as we show in Figure \ref{fig:CMD2013}\footnote{Please note that these mass estimates are made with respect to evolutionary models of single stars that do not include eruptive events and should be interpreted with caution.  We therefore do not expect that these accurately reflect the true initial mass of AT~2019krl or its actual evolutionary state, and it should not be surprising that observations at different epochs during an eruption may yield different mass estimates.  These are only meant to illustrate the equivalent mass of a single star that might have the same luminosity.}.   Even with lower amounts of extinction, 1.5 $< A_{V} <$ 2.0 mag, AT~2019krl would have a value safely within the expected masses of LBVs  \citep{2004ApJ...615..475S,2019MNRAS.488.1760S}.

\subsection{Comparison to Mergers}

Often referred to as red novae or LRNe, low-mass or intermediate-mass merger candidates can span a wide range of peak magnitudes and progenitor masses, yet may show a similar set of observational signatures \citep{2014MNRAS.443.1319K,Pastorello2019}. Merger candidates typically exhibit an initial peak in their optical light curve, followed by a secondary peak at some later date. Early-time spectra exhibit a blue continuum with narrow (100--300\,km\,s$^{-1}$) Balmer emission which fades with time as the spectra redden and cool, until finally molecular absorption lines appear and dominate the spectra a few months after maximum brightness.

In Figure \ref{fig:LCcomp} we compare the light curve of AT~2019krl to those of the massive star merger candidates M101-OT \citep{2017ApJ...834..107B} and SNHunt248 \citep{2015A&A...581L...4K,2015MNRAS.447.1922M,2018MNRAS.473.3765M}.   It is still unclear what mechanism is responsible for the multiple light-curve peaks; however, options include a common envelope (CE) ejection for the first peak and a second peak created during the final binary merger, or the first peak being caused by the adiabatic cooling of a CE event, while the second is from CSM interaction with the mass loss during inspiral \citep{2017MNRAS.471.3200M}. In the second scenario, viewing angle can easily change the observational signatures of the mergers. \citet{2020arXiv201014490S} estimates an up to 2 mag difference in brightness between equatorial observers and those viewing from the polar direction.
 
 It is possible that AT~2019krl had a double-peaked light curve missed by our sparse post-eruption observations. If so, the overall shape of the light curve fits those of merger candidates, with an absolute luminosity of the progenitor and the outburst being consistent with the class.  In particular, the color and temperature of AT~2019krl is quite similar to that of M101 OT2015-1, which was likely a YSG with $T_{\rm eff}$ = 7000\,K \citep{2017ApJ...834..107B}.  The progenitor mass and luminosity was quite a bit higher (18 M$_{\sun}$ and log($L$/L$_{\sun}$) = 4.9) for the M101 transient, but by adopting a moderately larger extinction, AT~2019krl could have a similar mass (Figure \ref{fig:CMD2013}). Of course, additional luminosity may come from the inspiral itself; therefore, the mass of the progenitor would be overestimated. The outflow speed implied from the P~Cygni absorption of 155\,km\,s$^{-1}$ in the H$\alpha$ emission is consistent with mass-loss speeds from the outer Lagrange point in stellar mergers \citep{2016MNRAS.461.2527P}, and is similar to the 150--200\,km\,s$^{-1}$ P~Cygni absorption seen in the light echoes of $\eta$ Carinae, which has been attributed to an outflow caused by the inspiral phase before a merger \citep{2018MNRAS.480.1466S}. 
 
One glaring discrepancy between AT~2019krl and merger candidates arises in the spectroscopic evolution. The H$\alpha$ emission is present and strong at all times in AT~2019krl, while in merger candidates it often fades after peak and may reemerge at late times. LRNe also lack the \ion{Ca}{2} NIR and [\ion{Ca}{2}] emission that we see in AT~2019krl.  Most notably, the molecular bands that form in merger spectra after $\sim 100$ days are not seen in AT~2019krl.  The lack of molecular lines is not unexpected, as our last photometric observation in 2019 December shows a transient with a temperature of at least 5250\,K, still too warm for the creation of molecular lines. 

A complication is that, in principle, mergers can occur across a wide range of initial masses, and mergers in more massive stars might not look the same as lower-mass mergers.  
Moreover, some individual LBV giant eruptions, including the prototypical case of $\eta$ Car, have been proposed as massive-star merger events \citep{2018MNRAS.480.1466S}, and mergers and mass gainers have been invoked to explain evolutionary considerations for LBVs more generally \citep{2014ApJ...796..121J,2015MNRAS.447..598S,2017MNRAS.472..591A}.  Therefore, the distinction between LBVs and low-mass merger events, such as V1309~Sco and LRNe, might arise simply from a continuum of different initial masses \citep{SmithNGC4490}, rather than distinctly different mechanisms.

\section{Conclusions} \label{sec:conclusions}

AT~2019krl clearly had a luminous and blue progenitor with no previous outbursts detected in the archival \textit{HST} and LBT images during the 16\,yr before the event. Observationally, it is consistent with known examples of giant LBV eruptions and SN~2008S-like objects with strong H$\alpha$, [\ion{Ca}{2}], and \ion{Ca}{2} NIR triplet emission and an estimated peak absolute magnitude between $-13$ and $-14$, yet it does not match a single class exactly.

The estimated mass of the directly-detected progenitor without any local extinction places the star in a mass regime of at least $\sim$ 13 M$_{\sun}$.  This is a lower limit because a modest increase in the adopted extinction correction could easily move the progenitor to higher masses.  Importantly, this moderately massive BSG progenitor is detected, despite the fact that the surrounding stellar population seems to indicate an older age and lower turnoff mass of only 8\,M$_{\odot}$.

We propose a scenario wherein AT~2019krl was the eruption of a BSG in a dense disk or toroidal CSM that was observed nearly pole-on.  A pole-on view of an object in a dusty torus is needed to simultaneously account for the presence of a strong IR excess and a seemingly contradictory lack of line-of-sight  extinction. This scenario could arise from binary interaction and a high-mass merger that resembled a giant LBV outburst.

Combined with what appears to be a fairly low-extinction environment, AT~2019krl may provide a link between SN~2008S-like transients and those occurring from unobscured progenitors, since similar observational properties of transient events seem to be occurring from very different progenitor types.
Deep UV-to-NIR late-time observations with very large ground-based telescopes, \textit{HST}, or \textit{JWST} will allow us to determine if indeed we have a hot luminous star cloaked in a massive dust shell created during the eruption, and if there is both a terminal and nonterminal eruption scenario that can create a very similar transient event.

\acknowledgments

We thank the referees for comments that helped us improve this paper, as well as the staffs of the observatories where data were obtained.
J.E.A. and N.S. receive support from National Science Foundation (NSF) grant AST-1515559. Research by D.J.S. is supported by NSF grants AST-1821967, 182197, 1813708, 1813466, and 1908972, as well as by the Heising-Simons Foundation under grant \#2020-1864.   Research by S.V. is supported by NSF grant AST-1813176. K.K. gratefully acknowledges funding from the Deutsche Forschungsgemeinschaft (DFG, German Research Foundation) in the form of an Emmy Noether Research Group (grant number KR4598/2-1).  TS is supported by the J\'anos Bolyai Research Scholarship of the Hungarian Academy of Sciences, by the New National Excellence Program (UNKP-20-5) of the Ministry of Technology and Innovation of Hungary, by the GINOP-2-
3-2-15-2016-00033 project of the National Research, Development and Innovation Office of Hungary (NKFIH) funded by the European Union, and by NKFIH/OTKA FK-134432 grant.
A.V.F.'s group received funding from NASA/{\it HST} grants GO-14668, 
GO-15166, and AR-14295 from the Space Telescope Science Institute (STScI), as well as from the TABASGO
Foundation, the Christopher J. Redlich Fund, and the Miller Institute
for Basic Research in Science (U.C. Berkeley; A.V.F. is a Miller Senior Fellow). Support for {\it HST} program GO-15151 was provided by NASA through a grant from STScI. 

Some of the data reported herein were obtained at the MMT Observatory, a joint facility of the University of Arizona and the Smithsonian Institution.
Research at Lick
Observatory is partially supported by a generous gift from Google.
Based in part on observations made with the NASA/ESA {\it Hubble Space Telescope}, obtained from the Mikulski Archive for Space Telescopes (MAST) at STScI. STScI is operated by the Association of Universities for Research in Astronomy, Inc. under NASA contract NAS 5-26555. 
Support for MAST for non-{\it HST} data is provided by the NASA Office of Space Science via grant NNX13AC07G and by other grants and contracts.
This work is based in part on archival data obtained with the {\it Spitzer Space Telescope}, which is operated by the Jet Propulsion Laboratory, California Institute of Technology, under a contract with NASA. Based in part on observations obtained at the Southern Astrophysical Research (SOAR) telescope, which is a joint project of the Minist\'{e}rio da Ci\^{e}ncia, Tecnologia, Inova\c{c}\~{o}es e Comunica\c{c}\~{o}es (MCTIC) do Brasil, the U.S. National Optical Astronomy Observatory (NOAO), the University of North Carolina at Chapel Hill (UNC), and Michigan State University (MSU).   Based in part on observations collected at the European Organisation for Astronomical Research in the Southern Hemisphere under ESO programme ID 098.C-0484(A).

The LBT is an international collaboration among institutions in the United States, Italy, and Germany. LBT Corporation partners are The University of Arizona on behalf of the Arizona university system; Istituto Nazionale di Astrofisica, Italy; LBT Beteiligungsgesellschaft, Germany, representing the Max-Planck Society, the Astrophysical Institute Potsdam, and Heidelberg University; The Ohio State University, and The Research Corporation, on behalf of The University of Notre Dame, University of Minnesota and University of Virginia. 
Some of the data presented herein were obtained at the W. M. Keck Observatory, which is operated as a scientific partnership among the California Institute of Technology, the University of California, and NASA; the Observatory was made possible by the generous financial support of the W. M. Keck Foundation. The authors wish to recognize and acknowledge the very significant cultural role and reverence that the summit of Maunakea has always had within the indigenous Hawaiian community.  We are most fortunate to have the opportunity to conduct observations from this mountain. 

\vspace{5mm}
\facilities{HST(ACS,WFC3,WFPC2), SST(IRAC), MMT(Binospec,BCH), Lick(Kast), LBT(MODS,LBC), Keck(Deimos), SOAR(Goodman)}

\software{astropy \citep{2013A&A...558A..33A}, }

\newpage

\bibliographystyle{aasjournal}
\bibliography{ms}

\end{document}